# Modellierungskonzepte der Synergetik und Theorie der Selbstorganisation

Werner Ebeling und Andrea Scharnhorst

## 1. Was verstehen wir unter Selbstorganisation und Synergetik? – Ein Überblick

### 1.1. Begriffsgeschichte und Kontext

Der Begriff *Selbstorganisation* spielt im letzten halben Jahrhundert eine zentrale Rolle in den Natur und Sozialwissenschaften. Wurzeln der Selbstorganisation können möglicherweise bis zur Antike zurück verfolgt werden werden. Der Begriff findet sich nachweislich auch in der Philosophie die 19. Jahrhunderts, z.B. in Schellings Philosophie der Natur. Wir werden uns aber in diesem Beitrag auf das 20. und 21. Jahrhundert, und hier insbesondere auf Modelle und Konzepte beschränken, die in dem Paradigmawechsel in der Physik hin zur nichtlinearen Dynamik wurzeln. Dennoch möchten wir uns, bevor wir Modelle aus dieser physikalischen Perspektive vorstellen, dem Begriff der *Selbstorganisation* auf einer allgemeineren Ebene annehmen.

In der russischen Physikalische Enzyklopädie (Band 4, 1994) (Afrajmovich, Rabinovich 1994) finden wir die folgende Definition: “Selbstorganisation – eine sich spontane Herausbildung stabiler regulärer Strukturen in dissipativen Nichtgleichgewichtsmedien, die nicht einer äusseren organisierende Einwirkung bedarf.” Verwiesen wird hierbei auf Arbeiten von Ilya Prigogine in den 1960er Jahren. Weiter heisst es im Text: “Den Prozess der spontanen Herausbildung regulärer Strukturen nennt man auch den Prozess der Formbildung, und das entsprechende Wissenschaftsgebiet nennt man häufig Synergetik”, wobei hier auf Herman Haken verwiesen wird. Obwohl dieser Eintrag viele spezifische Begriffe, wie Dissipation oder Nichtgleichgewichtsmedien enthält, deren Bedeutung sich erst im Kanon physikalischer Theorienbildung erschliesst, finden wir auch Elemente eines allgemeinen Verständnisses von Selbstorganisation. Es handelt sich um Prozesse, die spontan stattfinden, und zu Strukturen führen, die nicht von aussen, sondern aus dem Systeminnern selbst gebildet werden.

*Selbstorganisation* ist ein Begriff der eng mit der Systemtheorie verbunden ist, und der klar zu der heutigen Komplexitätsforschung gehört (Ball 2004; Mainzer 1997, 1999). Der Begriff *Synergetik* wurde von Hermann Haken Anfang der 1970er Jahre geprägt, und wird inzwischen zur Bezeichnung eines interdisziplinären Wissensgebietes verwendet. Die im Springer Verlag seit 1977 publizierte Buchserie *Lecture Notes in Synergetics* umfasst inzwischen 104 Bände und deren Titel verweisen auf alle wichtige Methoden und Anwendungsbereiche dieses Gebietes. Der Begriff *Synergetics* steht nicht nur für das von Hermann Haken gegründete Gebiet, wie man der englischen Wikipedia entnehmen kann. Der visionäre Architekt und Systemtheoretiker Richard Buckminster Fuller (Buckminster Fuller 1975, 1979; Kouw et al. 2013) verstand unter *Synergetics* das empirische Studium von Systemen im Übergang, wobei der Nachdruck darauf gelegt wird, das sich das Systemverhaltens nicht aus einer isolierten Betrachtung von Systemkomponenten, etwa dem Verhalten von Personen, die teils Akteur und teils Beobachter sind, verstehen und vorhersagen lässt.[1] Buckminster Fuller, Heinz von Förster, Stuart Kauffman und viele andere gehören zu der „Familie“ oder besser der wissenschaftlichen Gemeinschaft, die von Ideen von *Selbstorganisation* und *Synergetik* inspiriert war und ist, und wesentlich zu ihrem Verständnis beigetragen hat.

Ein Blick in die Wikipedia verrät, dass sich die Rezeptionsgeschichte des Begriffes *Selbstorganisation* im englisch-sprachigen Raum von dem im deutschsprachigen Raum signifikant unterscheidet. Was dem zweisprachigen Leser der entsprechenden Wikipedia Einträge ins Auge fällt, sind Differenzen sowohl der bezüglich der Themen als auch der Personen. Dies lässt sich auch graphisch, unter Benutzung von *eyeplorer* darstellen, einer sogenannten graphischen Wissenssuchmaschine. In Abbildung 1 wird die unterschiedliche Schwerpunktsetzung sichtlich, und in dem vergrösserten Ausschnitt der genannten Personen, gibt es nur einen Namen, der in beiden Wikipedias auftaucht: Ilya Prigogine. Es ist für eine erste Begegnung mit diesem Ansatz immer hilfreich sich zu vergewissern, auf welche Wurzeln im Sinne wissenschaftlichen Disziplinen, sich die jeweiligen Ansätze stützen. Ein Computerwissenschaftler wird Ansätze der Selbst-organisation anders präsentieren als ein Physiker, Chemiker oder Philosoph.

[1] http://en.wikipedia.org/wiki/Synergetics_Fuller

Der Begriff der *Selbstorganisation* ist heute in vielen Wissensgebieten verbreitet, und die damit verbundene wissenschaftliche Literatur wächst noch immer. In Abbildung 2 sieht man deutlich die beginnende Verbreitung des Begriffes zum Ende der 1960er Jahre. Gezählt werden dabei Artikel, die dieses Begriff im Titel, Schlagwort, oder Abstrakt verwenden. Um diese Grafik richtig zu interpretieren, muss man wissen, dass auch der Umfang der Datenbank mit der Zeit wächst (von etwa 15000 Dokumenten für das Jahr 1900 zu über 2 Millionen Dokumenten am Beginn der 2000er Jahre). Auch benutzt nicht jeder Artikel, der dem Gebiet der *Selbstorganisation* oder *Synergetik* zugeordnet werden kann, auch diesen Begriff. Andere Begriffe sind im Laufe der Zeit hinzugekommen, zum Beispiel Chaosforschung, aber vor allem Komplexitätsforschung und komplexe Netzwerke. Das heisst, dass das Wissensgebiet um den Begriff der *Selbstorganisation* noch viel grösser ist.

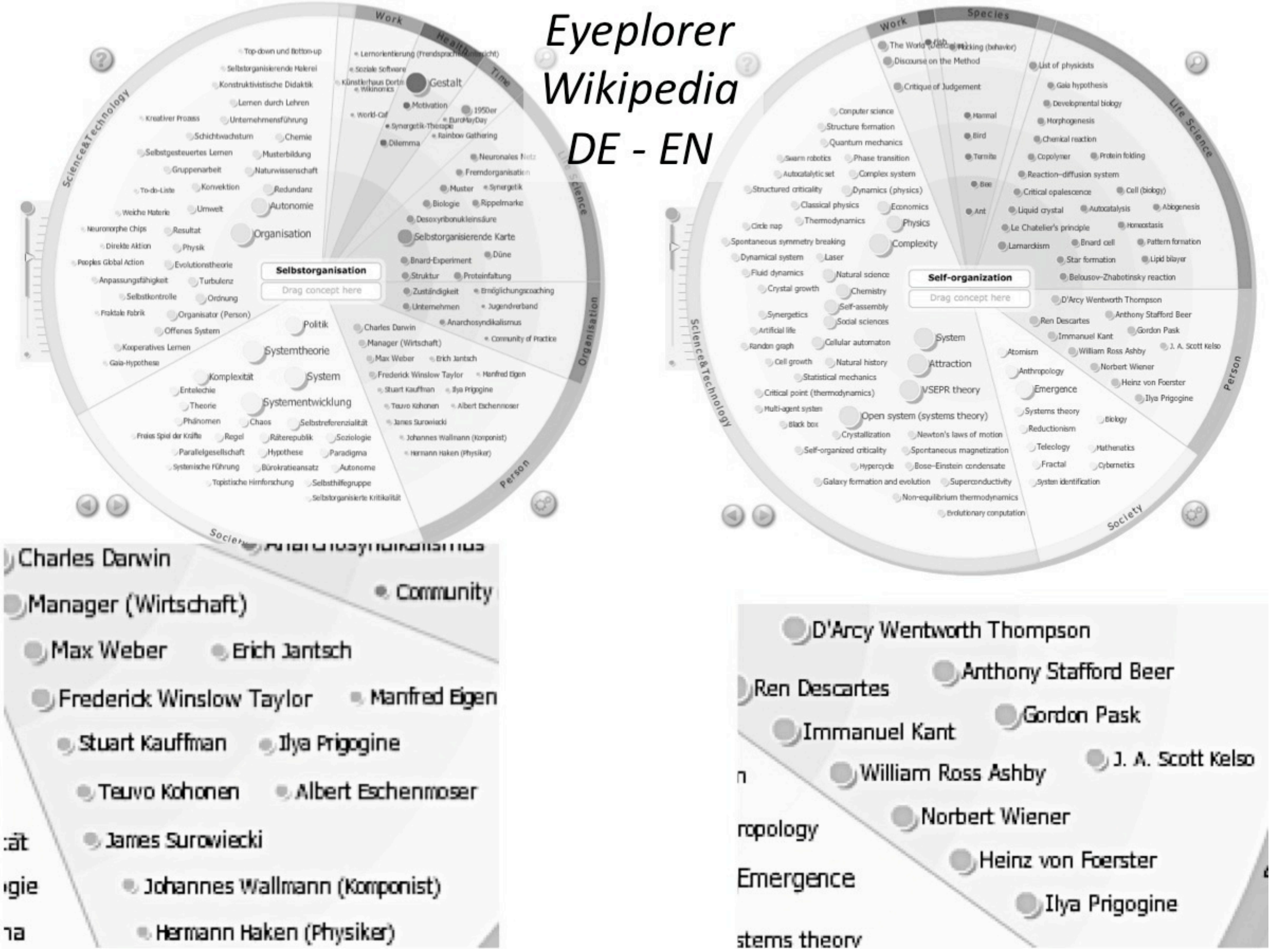

**Abbildung 1: Ausschnitte aus der graphischen Suchmachine *eyeplorer* für die Begriffe "Selbstorganisation" und "Self-organization" basierend auf Informationen aus der deutschen bzw. englischen Wikipedia. Vergrössert angezeigt sind die Namen von Personen, die in den jeweiligen Einträgen genannt werden.**

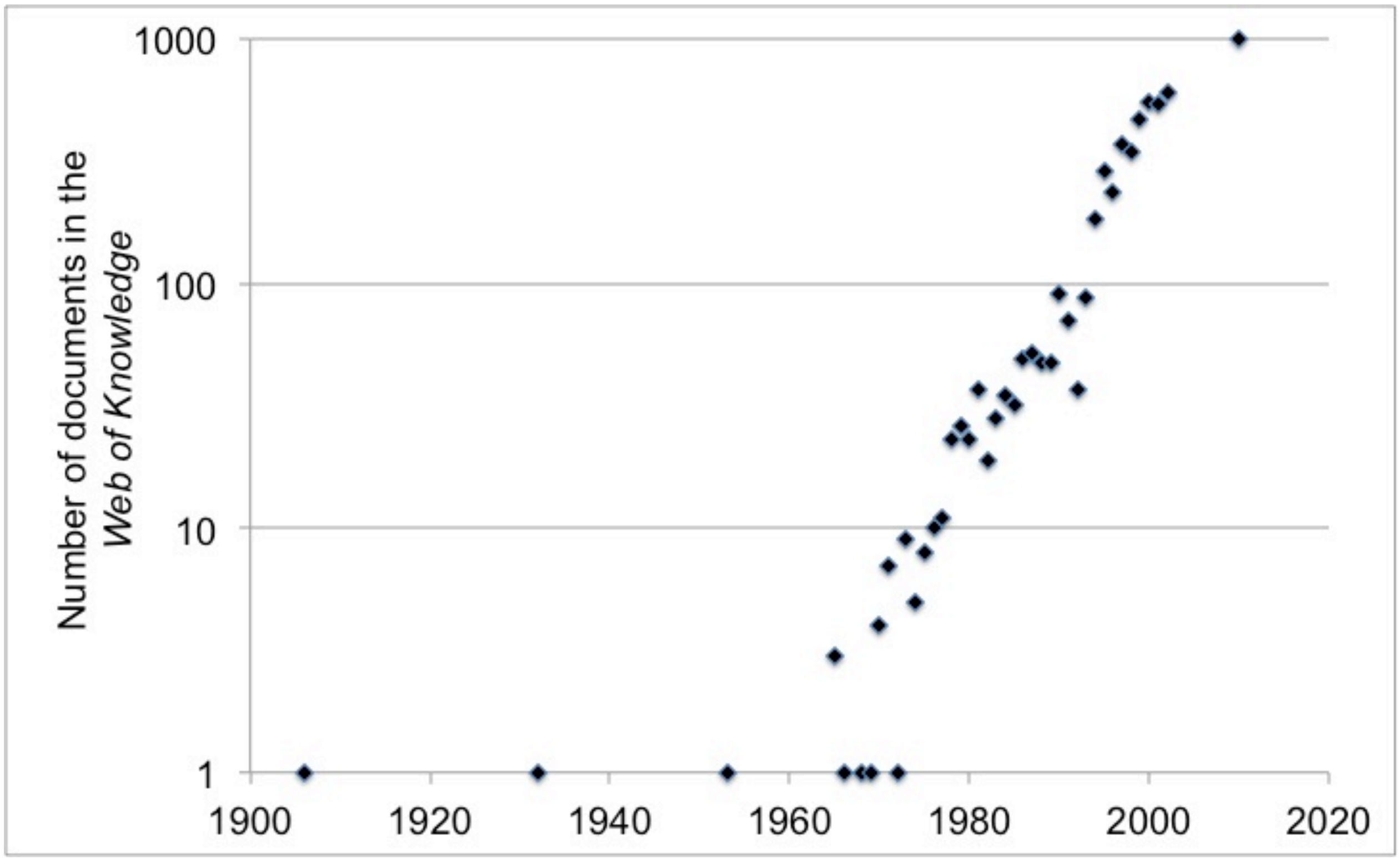

**Abbildung 2: Wachstum der Dokumente zur *Selbstorganisation* im *Web of Knowledge***

Bei dieser Verbreitung des Begriffes in, im Wesentlichen, allen Wissensgebieten, kann es auch nicht verwundern, dass um die Definition von Selbstorganisation noch immer gestritten wird. Der Informationswissenschaftler und Wissenschaftsphilosoph Klaus Fuchs-Kittowski etwa, hat die Begriffe Strukturbildung und Selbstorganisation einander entgegengestellt. Sein Vorschlag läuft darauf hinaus, im Bereich der Natur von „Strukturbildung" und nur in komplexen Systemen des gesellschaftlichen Bereiches von „Selbstorganisation" zu reden. Aber dieser durchaus sinnvolle Vorschlag hat sich nicht durchgesetzt.

In diesem Beitrag präsentieren wir Modellierungsansätze, deren Wurzeln vor allem in der statistischen, theoretischen Physik liegen. Wir werden später die rein physikalischen Anfänge näher beleuchten. In den 1970er Jahren sind es vor allem Forscher wie Eigen, Haken, Prigogine, in deren Arbeiten Grundprinzipien der Selbstorganisation formuliert und der allgemeinere Anspruch der entsprechenden mathematischen Modelle für Strukturbildung in Natur und Gesellschaft erhoben wurde. Diese Wurzeln teilt dieser Beitrag mit einem anderen im vorliegenden Handbuch. Dennoch können die resultierenden Modelle durchaus sehr verschieden sein, und es gibt darüber hinaus, auch keinen Grund nach einem einheitlichen Modell zu streben. Die Diversität von Modellansätzen hat durchaus eine erkenntnistheoretische Funktion.

Das Besondere der Physik der Selbstorganisation liegt darin, dass es sich dabei um allgemeine, abstrakte Theoriengefüge handelt, die aus Beobachtungen in der Natur entstanden sind, in ihrer weiteren Ausbreitung aber eine Allgemeingültigkeit beanspruchen, die jenseits der Dinglichkeit von Naturprozessen liegt. Das spiegelt sich auch in der folgenden Definition von *Synergetik*, als einem Wissensgebiet der Erforschung des Zusammenwirkens von Elementen gleich welcher Art, die innerhalb eines komplexen dynamischen Systems miteinander in Wechselwirkung treten. Dabei kann es sich um Moleküle, Zellen, Organismen oder Menschen handeln. Sie umfasst die allgemeingültigen Prinzipien und Gesetzmäßigkeiten des Zusammenwirkens von Elementen, die unabhängig von ihrer konkreten Realisierung in Physik, Chemie, Biologie, oder Soziologie sind und liefert eine einheitliche mathematische Beschreibung der räumlichen, zeitlichen und funktionalen Strukturen. Die spontane Bildung synergetischer Strukturen wird als Selbstorganisation bezeichnet. Das kollektive Verhalten gesellschaftlicher Gruppen ist eine Form der Selbstorganisation, wie an anderer Stelle in diesem Handbuch ausgeführt wurde (Haag und Müller, 2014). Bei Ebeling und Feistel finden wir die folgende Definition von Selbstorganisation, die auf einem ebenso allgemeinen Niveau operiert (Ebeling und Feistel, 1982, Feistel und Ebeling, 2011): „Der Begriff Selbstorganisation bezeichnet Prozesse, die weitab vom Gleichgewicht ablaufend, durch systemimmanente Triebkräfte zu komplexeren Strukturen führen“.

### 1.2. Die Aufnahme mathematischer Modelle aus der physikalischen Selbstorganisationsforschung in den Sozialwissenschaften

Es soll hier nicht unerwähnt bleiben, dass der Anspruch der Allgemeingültigkeit von Modellen und Konzepten der Synergetik und die Diffusion dieser Ideen in die Sozial- und Geisteswissenschaften auch auf Widerstand gestoßen ist, zum Teil noch stösst, und nicht ohne Kontroversen verlief. [2] “Menschen sind keine Atome” wurde der Synergetik entgegengehalten, und dass alle sozialen Prozesse *per definitio* nichtlinear sind. In der zum teil kontroversen Debatte, um Gesetzmässigkeiten in Sozialverhalten und Gesellschaftssystem, und deren Fassbarkeit in mathematischen Modellen wurde um eine neue Sprache, geteilt von Physikern, Computerwissenschaftlern, und Mathematikern auf der einen Seite und Soziologen und Philosophen auf der anderen Seite, gerungen. (Galison 1997) Nicht untypisch für die Diffusion neuer Ideen kam es dabei auch häufig zu Kommunikations- und Verständnisproblemen. Wie kann man die Eigenverantwortung und Selbstbestimmung der Menschen mit der Idee von Gesetzmässigkeiten im kollektiven Verhaltensspielfeld verbinden? Wird der Bedeutung der Kreativität handelnder Personen denn kein Abbruch getan, wenn man diese als blossen Faktor und in Form des so unscheinbaren und ungreifbaren Rauschens modelliert? Wie erfassen Modelle der Selbstorganisation sozialer Prozesse Planungs- und Leitungsvorgänge, als exogene Faktoren oder als Teil des Systems selbst? Wir werden auf einige dieser Fragen später noch zurück kommen. Wichtig ist uns an dieser Stelle zu betonen, das Synergetik und Selbstorganisation nicht *ein* Modell sozialer Prozesse bereitstellt. Vielmehr handelt es sich um ein Konzept zur Modellierung sozialer Prozesse, aus dem bereits eine Vielzahl von verschiedenen Modellen hervorgegangen ist. So wie man in der Architekturplanung mit Modellen verschiedener Art arbeitet (von gegenständlichen aus Pappmaché, über virtuell entworfene zur Testung von Statik,

[2] Siehe zum Beispiel Diskussionseinheit 8, Heft 4, Jahrgang 7, 1997 der Zeitschrift “Ethik und Sozialwissenschaften” zum Thema “Synergetik und Sozialwissenschaften” mit Beiträgen von 32 Autoren.

Akustik oder des Lichteinfalls, bis hin zu komplexen Entwurfsstudien, die u.U. einen Bogen spannen von der Geschichte des Ortes, seiner Kultur zu Formen möglichen zukünftigen Gebrauchs (Alkemade, 2006)), so bedarf auch die Modellierung gesellschaftlicher Prozesse einer *tool box,* mit ganz verschiedenen Denkwerkzeugen, von der Beobachtung, Befragung, hin zur Statistk und mathematischen Modellierung. (Merali, Allen 2011). Wir sind der Überzeugung, daß zu dieser tool box auch die Theorie der Selbstorganisation – Synergetik bereits wichtige Beitrage geliefert hat und noch liefern wird.

Trotz aller Geburtswehen dieser neuen, spezifischen Form der Modellierung komplexer sozialer Prozesse, hat die *wissenschaftliche Revolution der Selbstorganisation* (Krohn et al. 1990) von Anfang an eine grosse Faszination auf Geistes- und Sozialwissenschaftler ausgeübt. Die Diffusion von Ideen nahm dabei verschiedene Formen an, von der Übernahme von Konzepten, der Inspiration für neue empirische Studien, dem Entwurf von Simulationsprogrammen und analytischen mathematischen Modellen. Das heutige Gebiet der *computational sociology*[3], das ganz verschiedene mathematische Ansätze (von Agenten-basierter Modellierung bis hin zur Netzwerkanalyse) in sich vereint, zählt zu seinen Wurzeln auch Selbstorganisation. (Abbildung 3).

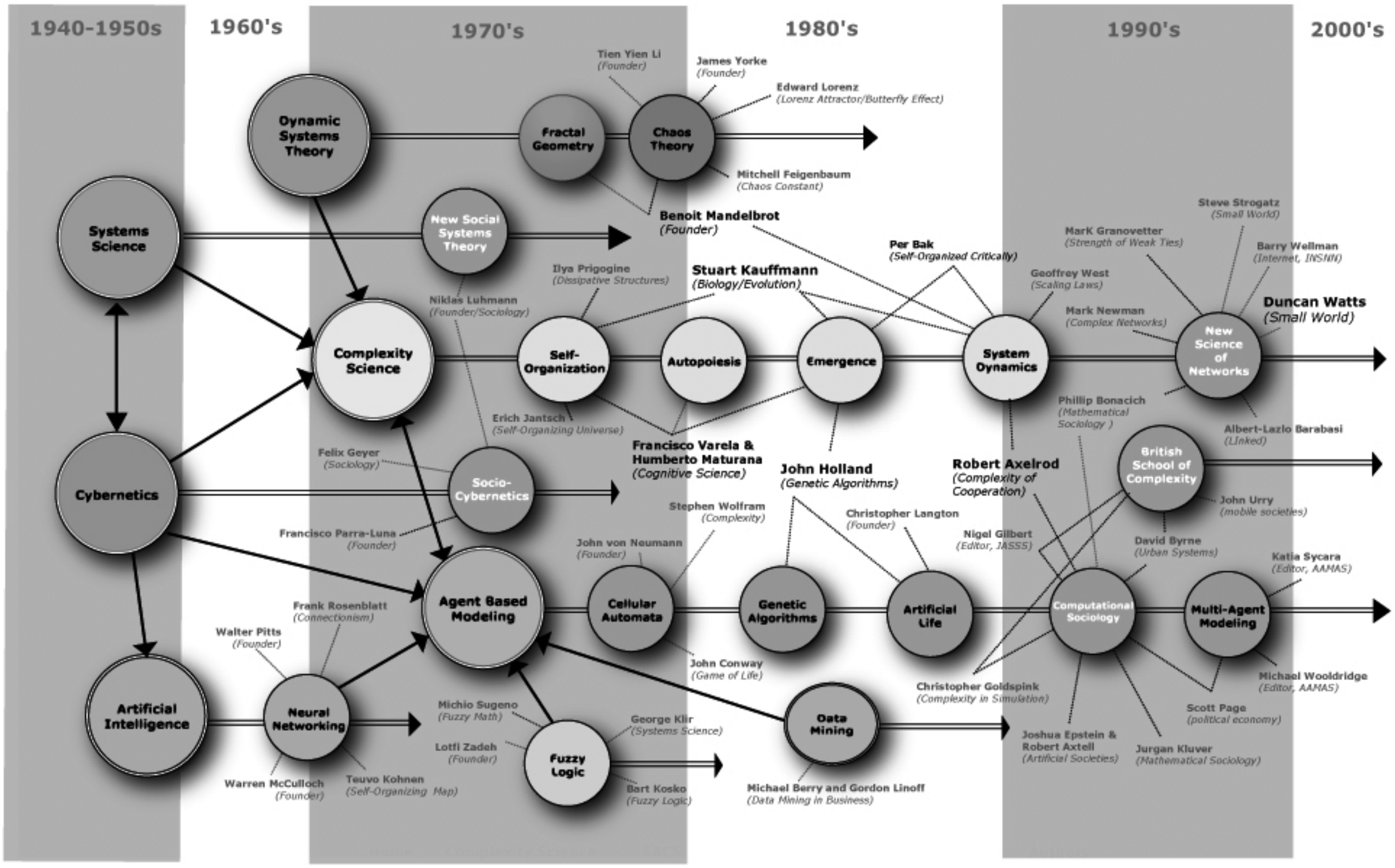

**Abbildung 3: Karte der Geschichte der Forschungsparadigmen und Wissenschaftler an der Schnittstelle von Soziologie und Komplexitätsforschung. Autor Brian Castellani. Quelle: http://en.wikipedia.org/wiki/File:Complexity-map-with-sociolo.png**

[3] http://en.wikipedia.org/wiki/Computational_sociology

Wir sind der Überzeugung, dass trotz oder gerade wegen des allgemeinen Charakters der Theorie der Selbstorganisation, sich der Geist und die Manier wissenschaftlichen Schließens und die Substanz bestimmter Modellansätze am besten erschliesst, wenn man sich der Geschichte ihrer Entstehung zuwendet.

# 2. Die Entwicklung der naturwissenschaftlichen Forschung zur Selbstorganisation und Synergetik

## 2.1. Schulen der physikalischen Selbstorganisationstheorien

Die Ursprünge der neueren physikalisch orientierten Theorie der Selbstorganisation und Synergetik liegen in den Arbeiten verschiedener wissenschaftlicher Schulen, die sich in den 60er und 70er Jahren des 20. Jahrhunderts herausgebildet haben. Es handelt sich dabei zunächst um die Brüsseler Schule um Ilya Prigogine, die etwa seit 1967 Arbeiten zur Selbstorganisation in thermodynamischen, hydrodynamischen, chemischen und biologischen Systemen durchführte (Prigogine 1967; Glansdorff & Prigogine, 1971) sowie um die Arbeiten der Stuttgarter Schule um Hermann Haken, die von der Lasertheorie ausgingen und das allgemeine Konzept der Synergetik begründeten (Haken, 1970; 1978; Haken & Graham, 1971). Diese beiden Schulen spielten eine prägende Rolle. Eine wichtige Rolle spielten auch die Arbeiten der Schule von Manfred Eigen in Göttingen zur Selbstorganisation und Evolution biologischer Makromoleküle (Eigen, 1971; Eigen und Schuster, 1977, 1979) und in Russland die Arbeiten von Yuri Klimontovich und seiner Schule in Moskau zur statistischen Physik offener Systeme (Klimontovich 1995).
Die moderne naturwissenschaftliche Begriffsbildung geht, wie bereits erwähnt, auf Manfred Eigen, Hermann Haken und Ilya Prigogine zurück. Einige Daten zu Leben und Werk der Pioniere: Ilya Prigogine, am 25.01.1917 in Moskau geboren, emigrierte mit seinen Eltern erst nach Berlin und dann nach Brüssel. Er besuchte die Schule in Berlin und studierte Physik und Chemie an der Universität Brüssel, wo er 1939 diplomierte und 1941 promovierte. Er erhielt seine Ausbildung im Geiste der berühmten belgischen Thermodynamik-Schule von DeDonder. Sein erstes, gemeinsam mit Defay verfasstes Buch (Defay, Prigogine 1946) ist den Grundlagen der Thermodynamik nach Gibbs und DeDonder gewidmet. Mit seinem zweiten Werk (Prigogine, 1947) schlägt er schon einen ganz originellen neuen Weg ein und wird so zu einem Begründer der Thermodynamik irreversibler Prozesse und der Theorie der Selbstorganisation, der er sich besonders in den 60er und 70er Jahren gemeinsam mit Glansdorff und Nicolis widmet. Seit 1951 war er als Professor für Physikalische Chemie an der Universität Brüssel tätig und seit 1977 auch als Leiter des „Center of Statistical Physics“ in Austin (Texas). Im Jahre 1977 wurde ihm der Nobelpreis für Chemie verliehen. Ilya Prigogine verstarb am 28.5.03 in Brüssel.
Manfred Eigen wurde am 09.05.1927 in Bochum geboren, Er studierte Physik und Chemie in Göttingen, wo er 1951 bei Arnold Eucken über schweres Wasser und elektrolytische Lösungen promovierte. Nach zwei Jahren der Zusammenarbeit mit Prof.

Ewald Wicke am Institut für Physikalische Chemie der Universität Göttingen wechselte er an das Max-Planck Institut für physikalische Chemie, das damals von Karl Friedrich Bonhoeffer geleitet wurde. Für seine Arbeiten über schnelle Reaktionen wurde Eigen 1967 mit dem Nobelpreis ausgezeichnet. Seit Ende der 60er Jahre befasste er sich mit der Selbstorganisation biologischer Makromoleküle, der Entstehung der Informationsverarbeitung und entwickelte mit Peter Schuster das Konzept hyperzyklischer Reaktionen (Eigen, 1971, Eigen und Schuster, 1977).

Hermann Haken wurde am 12.07.1927 in Leipzig geboren und gehört zu den führenden Vertretern der Nachkriegsgeneration der deutschen theoretischen Physik zählt. Nach dem Studium der Mathematik und Physik in Halle und Erlangen und der Promotion in Mathematik an der Universität Erlangen sowie Gastaufenthalten in Großbritannien und den USA wurde er 1960 auf einen Lehrstuhl für Theoretische Physik an der Universität Stuttgart berufen. Er arbeitet heute als emeritierter Professor des Lehrstuhls für theoretische Physik der Universität Stuttgart und gilt weltweit als Begründer der Synergetik. Zu seinen Arbeitsgebieten zählt die nichtlineare Optik, die Laserphysik, die Festkörperphysik und die statistische Physik. Hermann Haken schuf an seinem Stuttgarter Institut innerhalb kurzer Zeit ein internationales Zentrum der Lasertheorie. Bereits 2 Jahre nachdem 1960 der erste experimentelle Laser realisiert worden war, konnte Haken eine Theorie des Lasers vorstellen, die ihm große internationale Beachtung verschaffte. Die Interpretation des Laserprinzips als Selbstorganisation von Nichtgleichgewichts-Systemen führte Hermann Haken Ende der 1960er Jahre dann zur Entwicklung der Synergetik, einer neuen Wissenschaftsdisziplin welche die verschiedenen Ansätze aus der Thermodynamik, der Lasertheorie, und der nichtlinearen Dynamik zusammenfasst, und breite Anwendungen bis hin zu Modellen sozialer Prozesse anregt. (Haken, 1981, 1988)

Im folgenden wollen wir die Unterschiede zwischen den Ansätzen der Brüsseler und der Stuttgarter Schule herausarbeiten. Auf sophistische Überlegungen verzichtend, genügt es uns, grundlegende Lehrbücher der beiden Schulen aufzuschlagen (Glansdorff & Prigogine, 1971; Nicolis & Prigogine, 1977; Haken, 1978, 1988). Beim Studium dieser Lehrbücher erkennt man, dass der Brüsseler Ansatz im wesentlichen thermodynamisch orientiert ist und von Bilanzgleichungen ausgeht. Prigogine folgend ist das thermodynamische Nichtgleichgewicht die wichtigste Quelle von Ordnungsbildung in der Natur. Der Stuttgarter Ansatz ist mehr dynamisch - stochastisch bzw. systemtheoretisch orientiert und beginnt direkt mit nichtlinearen Differentialgleichungen und Langevin-Gleichungen. Man könnte auch sagen, der Brüsseler Ansatz legt den Schwerpunkt eher auf die makroskopischen, systemumfassenden Bedingungen, und deren Widerspieglung in makroskopisch zu definierenden Größen, während die Stuttgarter Schule sich viel stärker und differenzierter der Komplexität der Prozesse auf der Mikroebene zuwendet. Der Ansatz von Manfred Eigen, der etwa gleichzeitig entstanden ist, benutzt Elemente der beiden vorgenannten Schulen und wendet sie 1971 in einer viel beachteten Arbeit auf das Problem der molekularen Evolution (Eigen, 1971) und auf weitere Evolutionsprobleme an. In den in den 1980er Jahren erschienenen Ausgaben des Lehrbuches von Ebeling und Feistel (Ebeling, Feistel 1982; Ebeling, Engel, Feistel, 1990; Feistel, Ebeling, 2011), werden diese verschiedenen Ansätze vorgestellt, und dabei wird deutlich, dass es sich trotz unterschiedlicher mathematischer Sprache, um die verschiedenen Seiten einer Medaille handelt. Sowie sin der statistischen Physik Thermodynamik und statistische Mechanik zum Teil sich ergänzende Erklärungen ein und desselben Sachverhaltes sind, verhält es sich auch mit verschiedenen Ansätzen von Modellen der Selbstorganisation. Aber nicht einmal in der Physik, lässt sich jedes Modell in jedes andere Modell „übersetzen“. Für soziale Prozesse kommt das Fehlen des theoretischen und empirischen Kanons, dem Modelle minimal zu genügen haben, erschwerend hinzu.

Zur Geschichte der Physik der Selbstorganisation zurückkehrend, ist offenbar das Entstehen der statistischen Mechanik Ende des 19. Jahrhunderts, die eine mikroskopische, statistische Erklärung makroskopischer Phänomene liefert – man denke hier etwa an die Erklärung von Druck aus der Bewegung der Atome – eine grosse Anregung für die Erklärung andere Strukturen auf der makroskopischen Ebene gewesen. Als Pionier der naturwissenschaftlichen Untersuchungen zur Selbstorganisation in dieser Tradition gehört Hermann Helmholtz, der in seiner „Lehre von den Tonempfindungen“ (1877) wesentliche Grundlagen gelegt hat, die Rayleigh in seinem Werk „Theory of Sound“ (Rayleigh, Lindsay 1945) weiterführte und ausbaute. So sind die Ursprünge der Idee so eng mit der Theorie der Tonbildung und damit den physikalischen Grundlagen der Musik verbunden. Der dritte bedeutende Forscher des 19. Jahrhunderts war der bedeutende französische Mathematiker Henri Poincaré (Gray 2012), ihm verdanken wir die mathematische Grundlegung der Theorie. Es bleibt der Hinweis, dass auch Ludwig Boltzmann bereits ein lebhaftes Interesse für Selbstorganisation und Evolution gezeigt hat und wichtige qualitative Schlüsse ableitete (Broda, Gay 1983). Die Entwicklung im 20. Jahrhundert können wir hier nur durch eine Liste großer Namen und stichwortartige Kommentierung der Leistungen repräsentieren: Barkhausen entwickelte Anfang des 20. Jahrhunderts die Physik der selbsterregten Schwingungen (Barkhausen, 1907), die von Van der Pol ausgebaut wurde. Andronov begründete in den zwanziger und dreißiger Jahren eine bedeutende russische Schule der Theorie nichtlinearer Schwingungen. Hopf entwickelte in Leipzig die mathematische Theorie dazu, besonders in Bezug auf Bifurkationen höherdimensionaler Systeme. Schrödinger hat in seinem berühmten Buch „What is life?“ (Schrödinger, 1947) die physikalischen Grundlagen für Prozesse der Ordnungsbildung in großer Klarheit herausgearbeitet. Seine qualitativen Ansätze werden kurze Zeit später von Prigogine in eine quantitative Theorie umgesetzt (Prigogine, 1947). Mehr qualitativ sind wieder die Ansätze des österreichischen Biologen Bertalanffy, der Selbstorganisation der lebenden Materie beschreibt (Bertalanffy, 1953). Den bedeutenden russischen Forschern Kolmogorov und Bogoljubov verdanken wir wichtige Beiträge zu den mathematischen Grundlagen der Theorie (Bogoljubov, Mitropol'skij, 1961; American Mathematical Society, 2000). Lorenz studierte Anfang der 60er Jahre Selbstorganisation bei meteorologischen Prozessen und entwickelte grundlegende Ideen zum Begriff des Chaos. (Lorenz, 1993) Ende der 60er greift Prigogine mit Glansdorff und Nicolis erneut das Problem der Prozesse weitab vom thermodynamischen Gleichgewicht auf und analysiert die vorliegenden Experimente, wie die Belousov-Zabotinsky-Reaktion. Im Resultat entstehen die Grundlagen der modernen Theorie der Selbstorganisation (Glansdorff & Prigogine, 1971; Nicolis & Prigogine, 1977). Klimontovich in Moskau formuliert die statistische Physik der offenen Systeme (Klimontovich, 1995) Auch die mathematische Analyse wird weiter verfolgt, wobei besonders die Leistungen von Shilnikov, Smale, Arnold, Moser, Sinai, Ruelle und Takens hervorzuheben sind. Der Siegeszug einer Physik der Selbstorganisation beginnt.

## 2.2. Beispiele für Selbstorganisation in der Natur und Prinzipien der Theorie der Selbstorganisation

In diesem Kapitel gehen wir den folgenden Fragen nach: Wie manifestieren sich Phänomene der Selbstorganisation in der Natur? Welche empirischen Beobachtungen standen am Beginn der Theoriebildung? Wieso erfordert die Betrachtung selbst-organisierender Prozesse einen Paradigmenwechsel in der Physik? Welche erkenntnistheoretischen Einsichten innerhalb der Physik sind verantwortlich für ihre rapide Verbreitung ausserhalb der Physik, für Anwendungen des Konzepts der Selbstorganisation auf soziale Phänomene?
Die ersten Beobachtungen von Selbstorganisation wurden in Strömungen von Gasen und Flüssigkeiten gemacht. Um die Wende vom 18. zum 19. Jahrhundert beobachtete Benard, dass sich in einer Flüssigkeitsschicht, die von unten erhitzt wird, schöne hexagonal Strömungszellen ausbilden. Diese Strömungszellen werden generiert durch vertikale Temperaturgradienten, die einen kritischen Wert überschreiten müssen. Bei sehr großen Werten der Temperaturdifferenz zwischen unten und oben, gehen die regulären Zellen (auch Benard-Effekt genannt) bei einem kritischen Wert des Gradienten in Turbulenz über. Es erhob sich die Frage, ob die selbstorganisierten regulären Strömungen hier durch ungeordnete (turbulente) Strömungen abgelöst werden. Das entsprach lange Zeit der traditionellen Auffassung. Der russische Forscher Klimontovich, einer der Begründer der statistischen Theorie der Selbstorganisation entwickelte, unterstützt durch Prigogine, eine gegenteilige Auffassung: Turbulente Strömungen sind nicht weniger sondern mehr geordnet als die auffällig geometrischen Strukturen. Es handelt sich bei Turbulenz um eine besonders hochentwickelte komplexe Form der Strukturbildung; in einer turbulenten Strömung bewegen sich Milliarden von Molekülen in kohärenter Weise, die Molekülbewegungen sind hoch korreliert (Klimontovich, 1995).

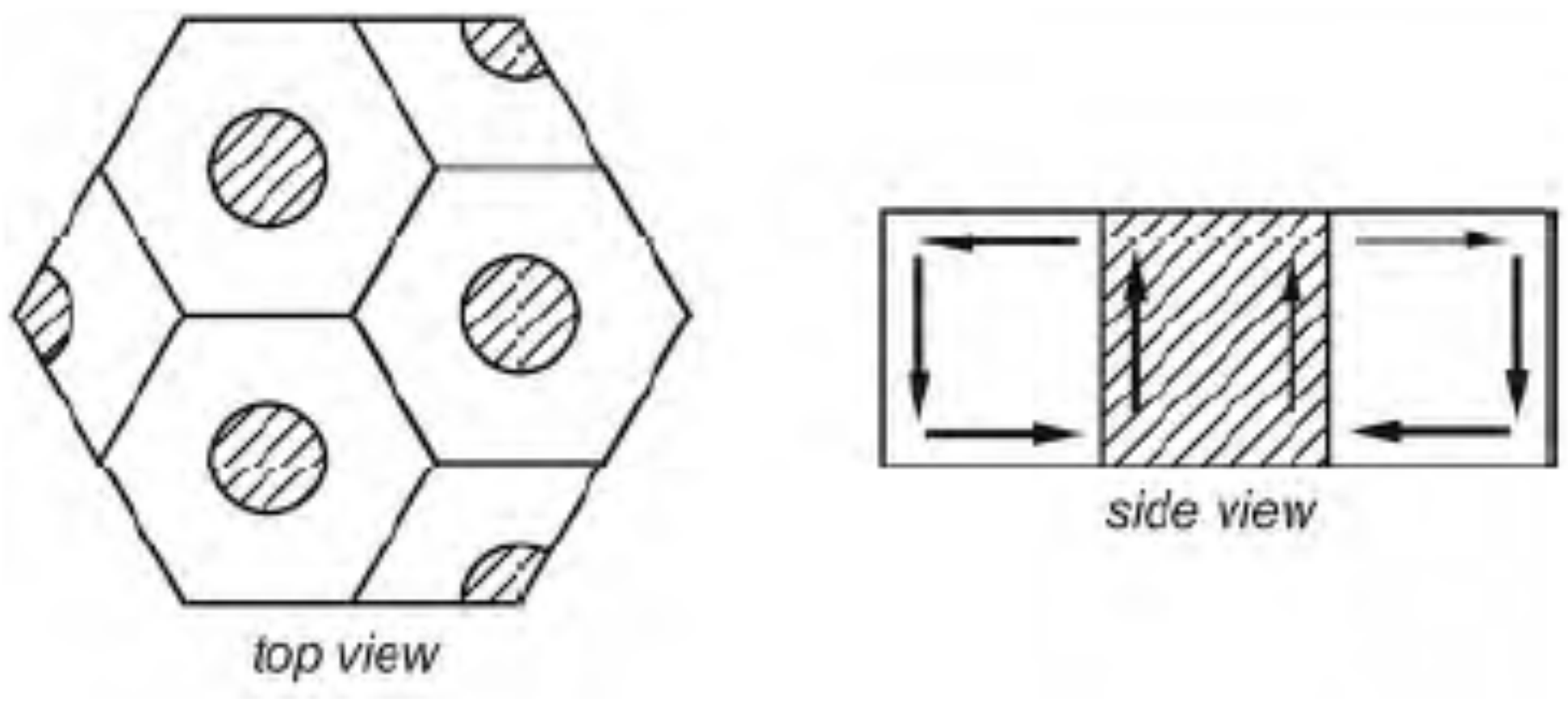

**Abbildung 4: Benard-zellen: Selbstorganisierte Flüssigkeitsstrukturen**

Ein anderes Beispiel für Selbstorganisation, in diesem Fall bei chemischen Prozessen, sind die bei der Belousov-Zhabotinsky-Reaktion auftretenden Muster und Wellen. Auch die noch älteren sogenannten Liesegangringe, die beobachtet werden, wenn bestimmte Chemikalien auf Filterpapier getropft werden, gehören zu den empirischen Befunden von Selbstorganisation. Ein weiteres Beispiel für eine physikalisch-chemisch und technisch relevante Selbstorganisation sind die elektrochemisch induzierten hydrodynamische Muster auf Elektroden (Plath 1989).

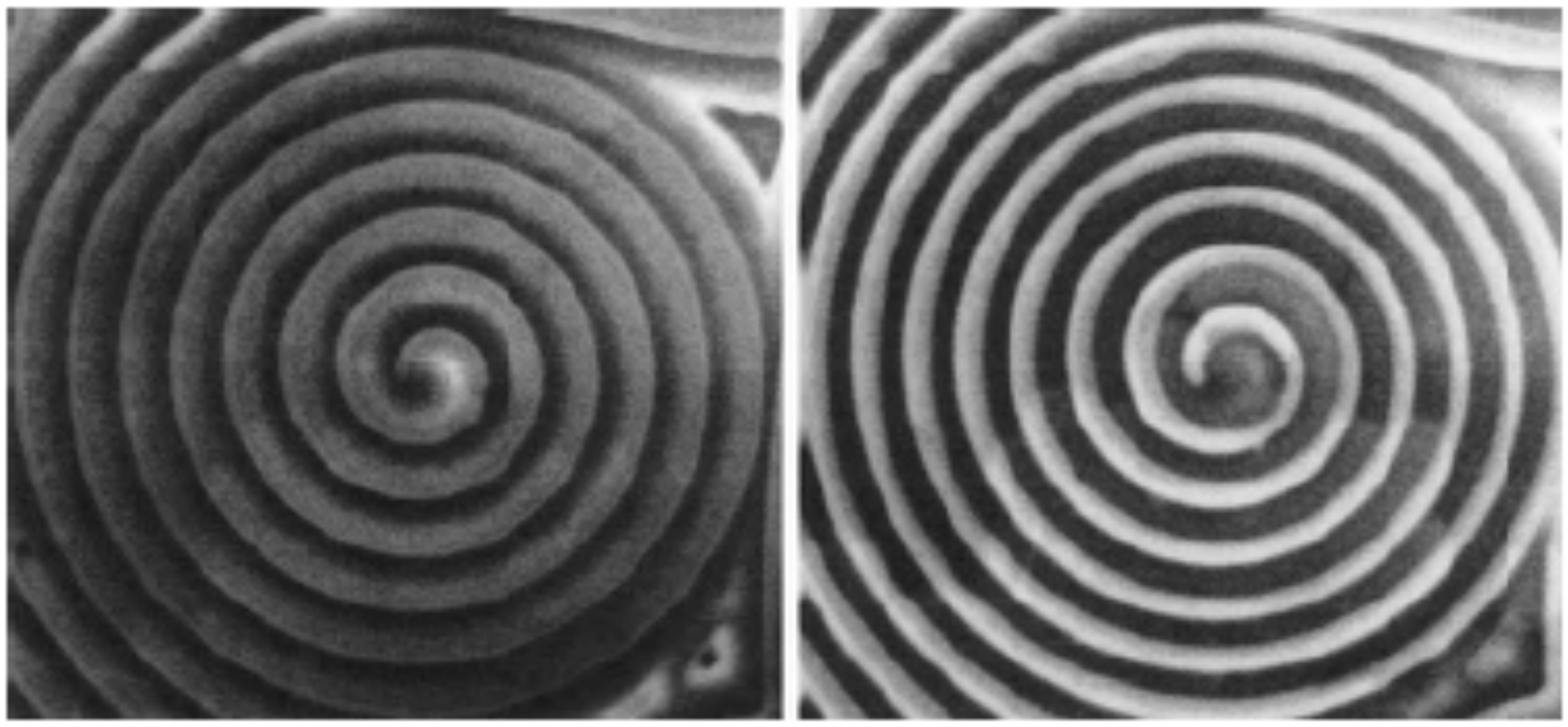

**Abbildung 5 Spiralstrukturen bei der Belousov – Zhabotinsky - Reaktion**

Die Beispiele, auf die wir eben kurz verwiesen, stehen für zwei verschiedene Phänomena in der Wissenschaftsdynamik um die Theorie der Selbstorganisation. Empirische Belege von Selbstorganisationsphänomenen finden sich weit zurück in der Wissenschaftsgeschichte, aber wurden in früheren Zeiten eher als Kuriosa behandelt, die ausserhalb des Erklärlichen stehen (etwa die Liesegangringe). Das Interesse an solchen Phänomenen wächst parallel mit der Entwicklung theoretischer Denkmöglichkeiten ihrer Erklärung. Aber noch erscheint es, als ob die spontane Strukturbildung ausserhalb des Bereiches liegt, für den das physikalische Theoriengebäude Erklärungen bereitstellt. Manchmal kommt es dabei auch zu einem: es kann nicht sein, was nicht sein darf Effekt. Im Fall der Belousov Zhabotinsky Reaktion etwa wird eine Arbeit abgelehnt, weil sie Phänomene beschreibt, die nach Massgabe der Referees so in der Natur gar nicht vorkommen können (Poleshchuk 1984, Pechenkin, A. 2009). In dem Masse wie das Verständnis von spontaner Strukturbildung wächst, schärft sich der Blick für Phänomene, die unter diese Kategorie fallen. Spontane Strukturbildungen werden von Exotika zur Norm beobachtbarer Phänomene, und noch in scheinbarer Unordnung – wie im Beispiel der Turbulenz – treten komplexe Ordnungen hervor. Ähnlich Prozesse der Ideenadaption lassen sich auch für die Diffusion der Konzepte der Selbstorganisation in den Sozialwissenschaften finden: der Werdegang von der Ausnahme zur Norm; eine Schärfung des empirischen Blickes für solche Phänomene; und die Entstehung von Theoriegebäuden und Methoden ihrer Erklärung.

Der Erkenntnisgewinn aus der Physik liegt vor allem in der Erklärung makroskopischer Strukturen aus spezifischen mikroskopischen Wechselwirkungen. Damit verbunden ist ein Perspektivenwechsel vom Studium bestehender, statischer Strukturen zu dem Prozess ihres Werdens. Zeit und die Unumkehrbarkeit von Prozessen tritt in den Vordergrund. Die oben dargestellten Beobachtungen aus der Natur verallgemeinernd kann man feststellen: Chaotische molekulare Dynamik, die normalerweise zum Ausgleich, und damit zum Verschwinden aller Ordnung führt, kann unter speziellen Bedingungen durch Selbstorganisation zu geordneten kohärenten Strukturen führen. Selbstorganisation ist das Gegenstück zur vorherrschenden Tendenz zur Unordnung. Die letztere wird durch den zweiten Hauptsatz der Thermodynamik erklärt. Selbstorganisation kann nur verstanden werden, wenn das Paradigma des zweiten Hauptsatzes aufgehoben wird. Diese Aufhebung nahm in der Geschichte der physikalischen Selbstorganisation eine spezifische Form an. Der zweite Haptsatz der Thermodynamik ist nicht falsch, aber sein Gültigkeitsbereich wurde neu definiert. Der scheinbare Widerspruch einer Verletzung des zweiten Hauptsatzes, wird dadurch aufgehoben, daß der Austausch mit der Umgebung in die Bilanzen einbezogen wird. Systeme, die traditionell als geschlossen betrachtet werden, um überhaupt beschrieben werden zu können, werden durch offene Systeme ersetzt. Für diese Systeme steht Strukturbildung nicht im Widerspruch zum zweiten Hauptsatz, wenn Entropieexport statt findet. Entropieexport bedeutet in anderen Worten, daß ein Import wertvoller Energie (Strahlung, Wärme hoher Temperatur, mechanische oder elektrische Arbeit) stattfindet, und daß wertlose Energie (z.B. Wärme niedriger Temperatur) exportiert wird. Der Import hochwertiger Energie und der Export geringwertiger Energie ist eine „conditio sine qua non“ für Selbstorganisation. Eine gleichwertige Aussage ist, daß Entropieexport stattfindet, der es gestattet, die unvermeidliche Entropieproduktion im Innern eines Systems zu kompensieren. Das Paradebeispiel für ein selbstorganisierendes System ist unsere Erde, die über den Strahlungshaushalt (Photonenmühle Sonne-Erde) die Fähigkeit zum Entropieexport besitzt (Ebeling und Feistel, 1982, Festel und Ebeling, 2011).
Was sind die Implikationen solcher Überlegungen für die Modellierung sozialer Systeme. Es ist nicht einfach Grössen, wie Entropie für soziale Systeme zu definieren, dennoch sind soziale Systeme auf jeden Fall offen, in dem obigen Sinne von Offenheit. Das Verhalten von Menschen bestimmt ihre Strukturen, und als biologische, lebende Wesen sind Menschen ‚offene’ Systeme. Aber Offenheit lässt sich auch in einem abstrakteren Sinne definieren. Das ökonomische System beruht auf der Produktion und dem Austausch von Waren, bei deren Herstellung nicht nur Stoffumwandlungen, sondern auch Wertzuwachs im Sinne menschlicher Arbeit auftritt. Das System der Wissenschaft, ein anderes Beispiel eines sozialen Systems, lebt von dem ständigen Zustrom ‚neuer’ ausgebildeter Wissenschaftler, und deren ständiger Ideenproduktion. Bei der Suche nach der Bedeutung von Einsichten aus der *Physik* der Selbstorganisation für die Selbstorganisation sozialer Prozesse, kommt es nicht auf eine mechanische Übersetzung der Grundbegriffe an. Physikalische Gesetze definieren den äusseren Bereich von möglichen Prozessen in Natur und Gesellschaft. Kein sozialer Prozess kann ein physikalisches Gesetz verletzen, aber physikalische Gesetze sind auch nicht in der Lage soziale Prozesse vollständig zu erklären. Physikalische, chemische, biologische und soziale Gesetzmässigkeiten bilden ineinander geschachtelte Trichter, die jeweils die Randbedingungen festlegen unter denen die immer komplexer werdenden Systeme operieren (Maturana, Varela, 1987). Was bleibt von den Einsichten physikalischer Selbstorganisation ist eine Orientierung auf Bilanzen, Prozesse, und ein erkenntnistheoretisches Herangehen, das nach Beschreibungsmöglichkeiten sucht, die aus der Komplexität des Geschehens, wesentliche Prozesse und Indikatoren bzw. Variablen zu extrahieren sucht. Dies wird besonders in der Synergetik sichtbar.

Nach der Definition von Hermann Haken ist die Synergetik die Wissenschaft von den komplexen, hierarchisch aufgebauten Systemen, die in der Lage sind, makroskopische räumliche, zeitliche oder funktionale Strukturen zu erzeugen. Es geht darum zu verstehen, wie das Zusammenwirken von vielen Teilsystemen, neue Strukturen und Funktionen auf der makroskopischen Skala erzeugen können. Das Prinzip der Ordnungsparameter[4] besagt, dass das Verhalten, also die Dynamik der Systemteile eines komplexen Gesamtsystems durch einige wenige Ordnungsparameter bestimmt wird. Damit findet, verglichen zur Komplexität bei der Betrachtung eines Einzelsystems, eine erhebliche Informationskomprimierung statt. Denn zur Verhaltensbeschreibung des Gesamtsystems reicht es abhängig vom Ordnungsparameter-Raum, einige, wenige Gleichungen aufzustellen, die das Gesamtsystem beschreiben. Ein anderes grundlegendes Prinzip ist das Versklavungsprinzip. Es sagt kurz formuliert aus, daß unter den vielen dynamischen Variablen eine Hierarchie der Wichtigkeit besteht. In der Regel sind nur eine wenige Parameter wirklich relevant. Das Spektrum der synergetischen Strukturen reicht von  einfacher Ordnung bis zur Komplexität hochorganisierter Systeme. Zentral für die Synergetik ist der Begriff des dynamischen Chaos. Durch die Arbeiten von Poincare, Lorenz, Shilnikov wissen wir, daß die nichtlineare Dynamik komplizierte Trajektorien generieren kann. Eine Bedingung dafür ist die dynamische Instabilität.

Zusammenfassend stellen wir fest, daß die wichtigsten Vertreter der naturwissenschaftlichen Theorie der Selbstorganisation in ihren Anliegen wesentlich übereinstimmen, sie folgen aber im Detail verschiedenen Ansätzen. Dies äussert sich auch in den verwendeten mathematischen Ansätzen. Der Brüsseler Ansatz ist im wesentlichen thermodynamisch orientiert und geht von Bilanzgleichungen aus.
Der Stuttgarter Ansatz ist mehr system-theoretisch, dynamisch und stochastisch orientiert, er beginnt direkt mit nichtlinearen Differentialgleichungen und Langevin-Gleichungen. Der Ansatz von Manfred Eigen bezieht sich vorwiegen auf Evolutionsprobleme. Er  benutzt Elemente der beiden vorgenannten Schulen und wendet sie auf das Problem der präbiotischen Evolution an und verallgemeinert diese dann auf Evolutionsprobleme in anderen Systemen.

# 3. Die Anwendung der von Theorien der Selbstorganisation auf gesellschaftliche Prozesse

## 3.1. Die gegenwärtige Forschungslandschaft von  Modellen der

[4] Siehe http://de.wikipedia.org/wiki/Ordnungsparameter

## sozialer Selbstorganisation

Wenn man den Bereich der Physik, Chemie und Biologie verlässt, und sich den Gesellschaftswissenschaften zuwendet, trifft man häufig auf außerordentlich komplexe Strukturen. Wir sind bereits im Kapitel 1 kurz auf den Siegeszug von Ideen der Selbstorganisation in den Sozial- und Geisteswissenschaften eingegangen, und werden dies am Beginn dieses Kapitels weiter vertiefen. In den letzten Jahrzehnten haben sich haben sich an den Berührungsstellen zwischen quantitative orientierter Sozialforschung und theoretischer Physik eigenständige Gebiete entwickelt, die Methoden der Synergetik und physikalischer Selbstorganisationstheorienaus soziale Prozesse anwenden. Dazu gehört die sogenannte *Econophysics* (Mantegna, Stanley 2000) und die *Sociophysics* (Schweitzer, 1997; Helbing, 1997, Galam, 2012). Auch in der Philosophie, vor allem der Wissenschaftsphilosophie haben Modelle dieser Art unter dem Bezeichnung computational philosophy (Thaggard, 1988) Einzug gehalten. Die computational soiology hat vor allem durch eine spezifische Modellklasse, die agenten-basierte Modellierung Auftrieb bekommen. Agentmodelle beruhen auf der Definition von (Verhaltens-)Regeln und eigenen sich daher besondern für eine Operationalisierung empirischer, systematischer Beoabchtungen und Verhaltensregeln und Normen, wie sich in der sozialwissenschaftlichen Theorien üblicherweise formuliert werden (Troitzsch1990; Epstein, Axtell, 1997; Gilbert 2010). Während *Socio*- und *Econophysics*, wie bereits der Name andeutet sich methodisch und erkenntnistheoretisch klar als Teilgebiete der Physik begreifen, haben die anderen genannten Strömungen Wurzeln in verschiedenen Gebieten der Mathematik, Systemtheorie, Künstlicher Intelligenz, und beziehen sich in der Regel explizit auf Theorien der für die jeweilig betrachteten Phänomene relevanten geistes- und sozialwissenschaftlichen Disziplinen.

Für die Forschungsarbeiten, die sich noch stets als der Physik, oder besser der mathematischen Physik zugehörig empfinden, steht die Entwicklung von allgemeinen Methoden zur Modellierung sozialer Prozesse im Vordergrund. Die Mathematik ist entsprechend komplex und die in diesen Richtungen erzielten Ergebnisse und Einsichten verbreiten sich daher vorrangig unter Forschern, die der entsprechenden mathematischen Sprache mächtig sind. Für die Pioniere, die ihre Wurzeln in einer der Sozialwissenschaften, sei es Ökonomie oder Soziologie, haben, hat dagegen der Bezug neuer Modellansätze zu bestehenden Theorien und empirischen Studien in diesen Gebieten Vorrang. Im Endeffekt, wie von Galison beschrieben (Galison 1997), verlaufen Prozesse der Integration neuer Konzepte, neuer mathematischer Methoden und neuer empirischer Studien in den noch immer voneinander geschiedenen disziplinärer Strömungen häufig asynchron in Zeit und was die Abfolge der Integration neuer Ideen angeht. Dies wird in der Abbildung 3 in den verschiedenen Strömungen der *computational sociology* sichtbar. Um dies an einem weitere aktuellen Beispiel zu illustrieren, genügt es exemplarisch Arbeiten aus der gegenwärtigen Netzwerkforschung zu betrachten. Unterschiedliche Interessen und Forschungsschwerpunkte werden dabei sichtbar. (Pyka, Scharnhorst 2009) Während es für einen Sozialwissenschaftler etwa selbstverständlich ist, das soziale Netzwerke aus Knoten verschiedenster Art zusammen gesetzt sind und das Personen, zeitgleich in verschiedenen Netzwerken und darin in unterschiedlichen Rolle agieren können, bildet die mathematische, analytische Behandlung solcher Multiplexe (Baxter et al 2014) noch stets eine Herausforderung.

Zusammengefasst ergibt sich das Bild einer Forschungslandschaft, in der aus den Kernbereichen von Physik und Mathematik einerseits, und Sozial- und Geisteswissenschaften andererseits, Forscher mit unterschiedlichem akademischen Hintergrund an ähnlichen Problemen arbeiten, und sich sozusagen aufeinander zubewegen. Sie alle teilen ein epistemisches Grundgerüst, dessen Basisprinzipen sich folgermaßen definieren lassen:

- Soziale Systeme sind komplexe System und folgen Gesetzmäßigkeiten ähnlich zu anderen komplexen Systemen.
- Die Aktionen und Interaktionen von Akteuren in diesen System führen zur Entstehung von Strukturen auf der Systemebene.
- Ein Verständnis dieser Strukturen ist nur möglich, wenn die Prozesse ihres Werdens in die Erklärung aufgenommen werden. Zeit wird zur konstituierenden Dimension, und die Unumkehrbarkeit von Prozessen zum wesentlichen Merkmal.
- Es entsteht ein Kreislauf zwischen selbstbestimmtem individuellen Handeln; der spontanen Emergenz systemimmanenter Strukturen jenseits des einzelnen Individuums; und des Wirkens dieser systemweiten Bedingungen als Handlungsrahmen für das selbstbestimmte individuelle Handeln.
- (Mathematische) Modelle für die Aktionen und Interaktionen der Individuen und die resultierenden Strukturen gehören zur Klasse von Modellen für dynamische, nicht-lineare Prozesse in offenen Systemen.

Unterstützt von einer Diffusion von analytischen und computergestützten Methoden entstehen dabei Übergangsbereiche, in denen sich neue erkenntnistheoretische Normen und methodischen Grundfertigkeiten etablieren. Die gegenwärtige Situation ist vor allem von einer Koexistenz, teilweiser Redundanz bis hin zu gelegentlicher Unkenntnis paralleler Problemlösungsbestrebungen gekennzeichnet. Zum Teil ist diese Diversität von Modellansätzen nicht nur berechtigt sondern auch notwendig. Ebenso wie die Physik verschiedene mathematische Modelle für verschiedene Phänomene entwickelt hat, wäre es naiv, für die noch komplexeren sozialen Phänomene *ein* Modell oder einen Modellansatz zu erwarten. Andererseits ist es, angesichts der relativen *Jugend*, in der sich dieses Forschungsgebiet befindet, nach wie vor erforderlich den Überblick über die verschiedenen Modellansätze zu fördern und wo es angebracht ist, auch um einen wechselzeitigen Bezug, oder um eine Übersetzung dieser Modelle ineinander zu ringen; in anderen Worten um einen relative konsistenten Theorie-, Experiment und Methodenrahmen. Das ist das Anliegen dieses Handbuch und unser Beitrag ist dabei den Ansätzen einer bestimmten Richtung, der Ebelingschen Schule, gewidmet. Innerhalb der Physik basieren diese Konzepte auf einer Integration der Ideen der drei oben genannten Schulen (Feistel & Ebeling, 2011). Der Erklärungsanspruch des resultierenden Theoriegerüsts ist universell. Es handelt sich um allgemeine evolutionstheoretische Betrachtungen, die historisch in einer Linie mit Alfred Lotka's Programm einer Physik der Evolution stehen. (Lotka, 1911)

Wir werden im Folgenden die Grundzüge dieses Herangehens herausarbeiten, dann einen spezifischen Modellierungsansatz näher beleuchten; und im letzten Teil dieses Kapitels ein Evolutionsspiel vorstellen.

## 3.2. Innovation, Instabilität und universelle Gesetze in der Entstehung des Neuen

Wir verstehen sozioökonomische Prozesse als spezielle Evolutionsprozesse auf der Ebene der Gesellschaft. Ein Kennzeichen dieser Prozesse ist ihre Historizität, sie sind Ergebnis einer langen Folge von Teilprozessen. Wir verstehen solche Prozesse als praktisch unbegrenzte Folgen von Schritten der Selbstorganisation (Ebeling & Feistel, 1982, 1986; Ebeling, Engel & Feistel, 1990). Im folgenden werden wir uns auf Modellansätze konzentrieren, die genau diesen Übergangsbereich beschreiben: das Entstehen, Eindringen und mögliche Durchsetzen einer Neuerung gegenüber einer bereits etablierten Struktur. Zu Beginn gilt es aber einige Grundbegriffe zu klären. Was verstehen wir unter System und Struktur? Was sind die zentralen Objekte der Modellierung und wie verhalten sich die abstrakten Eigenschaften mathematischer Modelle von Systemdynamik zu den jeweils konkreten Phänomenen?

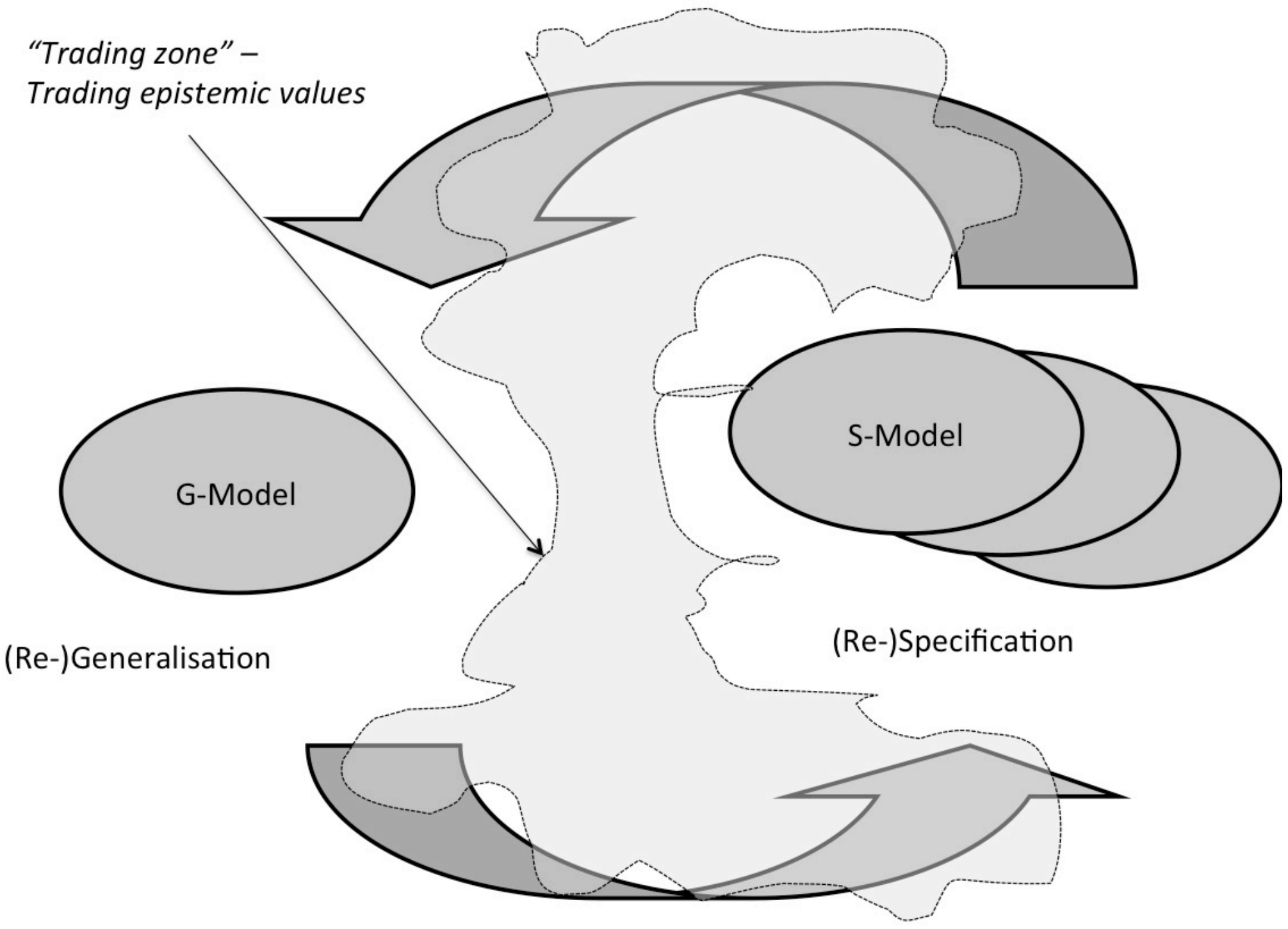

**Abbildung 6 Der Zyklus zwischen universeller, allgemeiner Modellbildung (general models) und der Anwendung auf konkrete Phänomene (specific models) (nach Lucio-Arias, Scharnhorst 2012)**

Auf der Ebene eines allgemeines Modellrahmens, und bevor wir eine konkrete Anwendung ins Auge fassen (zum Zyklus der Modellbildung zwischen Allgemeinem und Spezifischem siehe Abbildung 6), beschreibt der Begriff *Struktur* eine relative stabile Konfiguration bestehender Verhaltensmöglichkeiten. In der Physik spricht man dann auch von Zuständen eines Systems. Die Verhaltensmöglichkeiten selbst sind die Einheiten der Evolution, im selben Sinne wie biologische Populationen Einheiten der biologischen Evolution sind. Wir sprechen dann auch von Gruppen, Populationen oder

Sorten. Diese unterscheiden sich voneinander, und lassen sich entweder so voneinander abgrenzen, dass sie als disjunkte Gruppen beschrieben werden (diskrete Beschreibung) oder sie werden durch Merkmale beschrieben, die sich kontinuierlich verändern können (kontinuierliche Beschreibung). Welche Verhaltensmöglichkeiten (Gruppen) in einem System realisiert werden, erkennt man empirisch durch die Zuordnung von Akteuren (Elemente) zu den Verhaltensmöglichkeiten. Dieser mathematische Ansatz ist universell, und kann auf ganz verschiedene Situationen angewendet werden. Die *Kunst* der Modellbildung besteht nun darin, die Einbettung oder Re-spezifizierung so vorzunehmen, dass wir beobachtbare Phänomene beschreiben können, und neue Einsichten in ihr Zustandekommen gewinnen.

Wenn wir von einer Struktur sprechen, gehen wir davon aus, dass nicht alle möglichen Verhaltensweisen oder Gruppen gleichermassen im System vorkommen sondern, daß es gewissen Präferenzen gibt. Ein Beispiel ist die Dominanz einer Verhaltensweise.
In den Modellansätzen der Ebelingschen Schule zur Selbstorganisation sozialer Prozesse finden wir eine gewisse Dreiteilung von einerseits Struktur – auf der Makroebene; den Einheiten, die diese Struktur ausmachen – Gruppen auf der Mesoebene; und den Akteuren, die sich Gruppen zuordnen lassen und deren Verhalten, und Interaktion auf der Mikroebene die Dynamik im System ausmacht, und zur Entstehung der Strukturen beiträgt. Es ist dieselbe Dreiteilung, die wir im vorherigen Kapitel in den Modellen der physikalischen Selbstorganisation vorgestellt haben. In Tabelle 1 geben wir einen Überblick über mögliche *Übersetzungen* dieses allgemeinen Modellansatz für verschiedene Anwendungen.

**Tabelle 1: Verschiedene Anwendungen eines Modellansatzes zur Selbstorganisation und Evolution**

| Anwendungsbereiche/ Modellansatz | Wissenschaft | Wirtschaft | Wissensmanagement |
|---|---|---|---|
| System | Anzahl wissenschaftlicher Gebiete | Ein industrieller Sektor | Eine Unternehmen |
| Phänomen (Makroebene) | Wissenschaftliche Gebiete entstehen und vergehen. | Neue technologische Innovationen setzen sich durch, oder werden an einer Durchsetzung gehindert (*lock-in*) | Die Suche nach Problemlösungen in einer Gruppe von Personen mit unterschiedlichen Kompetenzen. |
| Gruppen (Mesoebene) | Wissenschaftliche Gebiete | Technologien, die in diesem Sektor benutzt werden | Unterschiedliche Problemlösungsvorschläge. |
| Akteure | Wissenschaftler | Firmen | Personen |
| (Mikroebene) | | | |
| Struktur | Die Verteilung von Wissenschaftler zwischen den bestehenden Gebieten. | Verteilung von Technologien über die Firmen. | Existierende Problemlösungen, und wie viele Gruppenmitglieder diesen anhängen. |
| Innovation | Ein neues Gebiet, das die bestehende Konfiguration – d.h. die Verteilung von Wissenschaftlern auf verschiedene wissenschaftliche Gebiete – entscheidend verändert. | Eine neue Technologie setzt lösst eine bestehende Technologie ab, auch gegen Widerstände *(Lock-in)*. | Die Gruppe findet eine neue Problemlösung. |
| Weiterführende Literatur | Bruckner et al. 1989, 1990; Vitanov, Ausloos 2012 | Bruckner et al. 1996, Saviotti 1996, Ebeling et al. 2001, Aigle et al. 2008, | Erpenbeck, Scharnhorst 2005, Erpenbeck et al. 2006 |

Die Existenz einer relativen stabilen Struktur hat dabei nichts damit zu tun, dass das System ein Gleichgewicht erreicht hat, in dem keine Veränderungen mehr stattfinden. In der Regel handelt es sich in sozialen Systemen um Fliessgleichgewichte (Bertalanffy 1953). D.h. obwohl sich relative stabile Verhältnisse herausgebildet haben, finden nach wie vor Prozesse statt, nur tragen diese am Ende zu einer Stabilisierung einer bestimmten Struktur bei.

Die Strukturen in sozialen Systemen lassen sich empirisch ebenso beobachten wie Strukturen in der Natur. Sie vertonen in manchen Darstellungsweisen sogar visuelle Ähnlichkeiten zu den Abbildungen aus Kapitel 3.

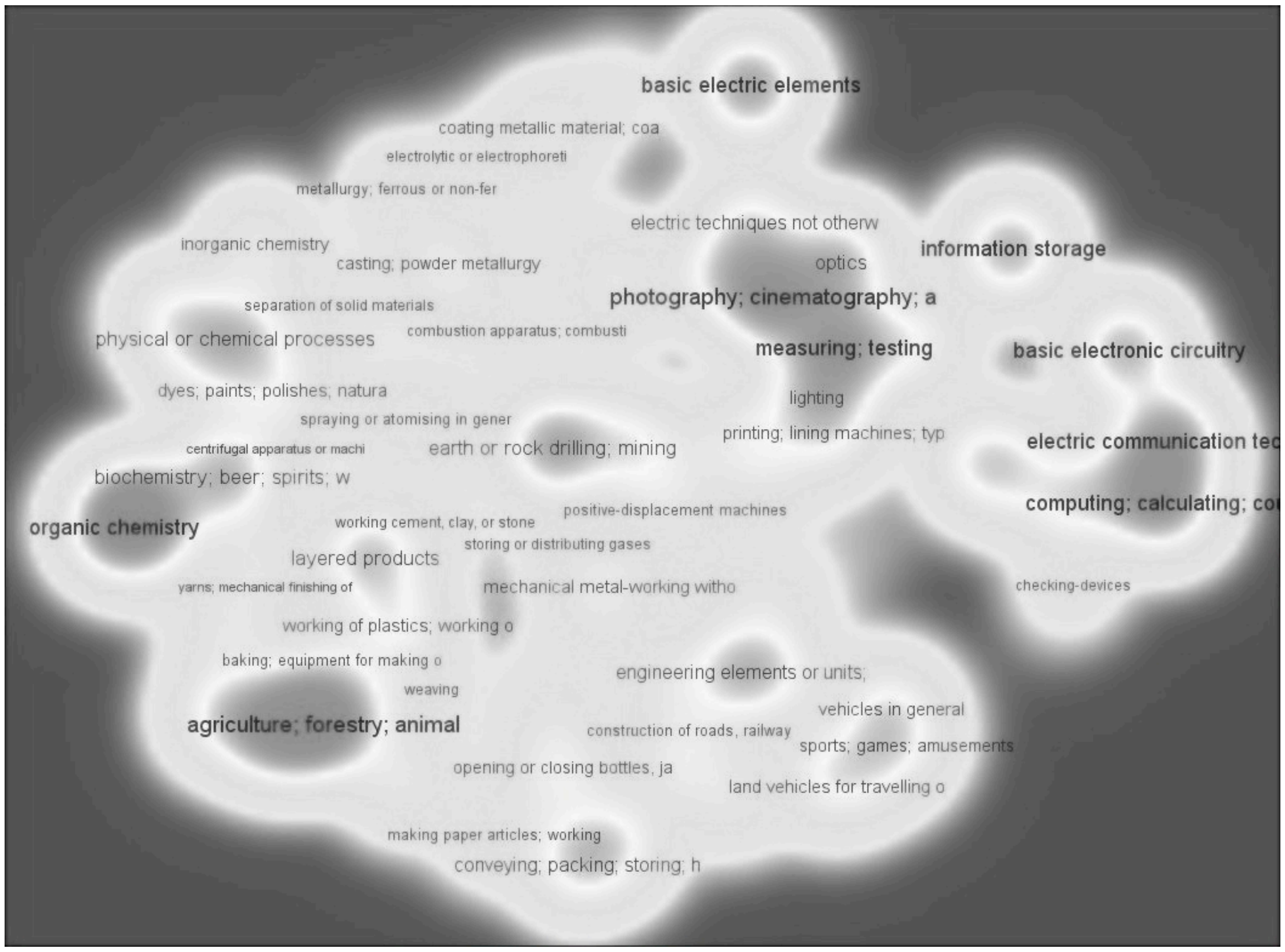

**Abbildung 7 Heat map[5] der wissenschaftlichen Themen in der Patentliteratur. Dargestellt ist eine Menge von US Patenten mit einem niederländischen Erfinder aus dem Jahr 2007 (Siehe Leydesdorff & Bornmann, 2012, pp. 1446 ff und http://www.leydesdorff.net/ipcmaps/ für weitere Erläuterungen).**

Wenn wir von Innovation sprechen, dann ist es die De- oder Instabilisierung genau dieses *status quo.* Denken wir etwa an ein Geflecht von Technologien, von denen eine, trotz der Existenz alternativer Technologien, noch immer dominant ist. Dies kann man gegenwärtig im Fall des Verbrennungsmotor für Autos beobachten (Aigle et al. 2008). Eine De-stabilisierung wäre das Brechen dieser Dominanz.

Im Fall des Wissenschaftssystems läßt sich ein gewisser *status quo,* eine stabile Struktur, auf der Ebene der Disziplinen beobachten. Diese verändern sich relativ langsam – die letzte große Umwälzung ist etwa das Aufkommen der Computerwissenschaften in den 50er Jahren. Von dieser relativen Konstanz wird in den so-genannten *base maps* der Wissenschaft (Börner et al. 2003; Börner 2010; Börner et al. 2012; Rafols et al. 2010) Gebrauch gemacht. Diese visuelle Darstellungen zeigen Grundkonfigurationen zwischen wissenschaftlichen Gebieten auf Basis von Zeitschriften, und den Verbindungen zwischen Referenzen (Zitaten) in den in den Zeitschriften publizierten Artikeln. Da sie über Jahrzehnte relativ unverändert sind, eignen sie sich als Bezugsrahmen für die Darstellung von Ausbreitungsphänomenen wissenschaftlicher Ideen.

Unabhängig von der spezifische Gestalt des Systems, der darin zu beobachtenden Struktur, des dominanten Regimes oder Verhaltensmuster, lassen sich auf der Ebene des allgemeinen Theorienrahmens Regeln definieren, unter denen ein Wechsel von

[5] Unter Benutzung von VosViewer http://www.vosviewer.com/ (Van Neck, Waltman 2010)

einer Struktur zu einer anderen möglich erscheint. Das Auftreten von Innovationen auf der Ebene des Gesamtsystems ist immer mit der Destabilisierung eines bestehenden und Restabilisierung eines neuen Zustandes in einem weiteren Selbstorganisationsschritt verbunden ist (Abbildung 8). Mit anderen Worten, strukturbildende Prozesse sind gleichzeitig immer auch strukturzerstörende Prozesse.

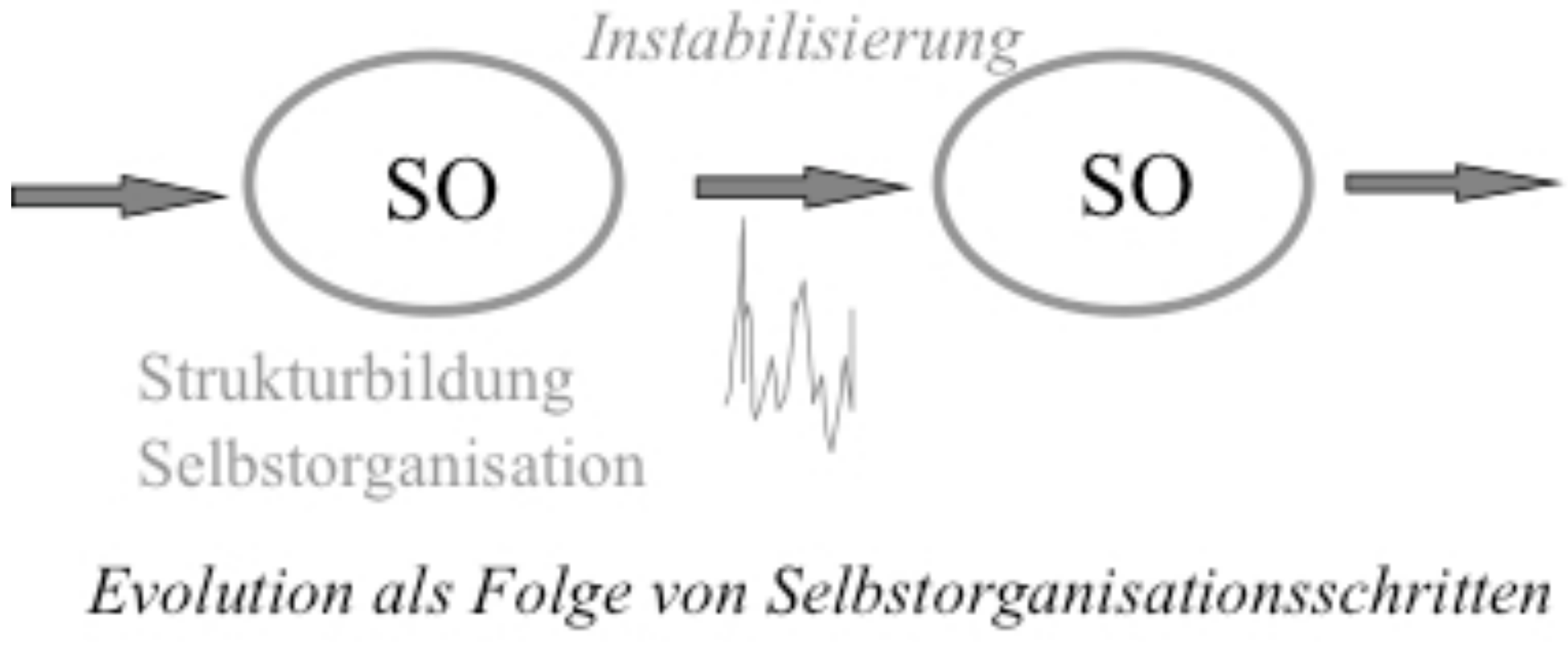

**Abbildung 8 Schema von Folgen von Prozessen der Selbstorganisation**

Was als Neues oder Innovation im breitesten Sinne angesehen wird, hängt entscheidend davon ab, wo die Grenzen des Systems gezogen werden. Betrachten wir als Beispiel die Wirtschaft, technologische Innovationen, und deren Nutzung durch Firmen. In einem industriellen Sektor etwa wird eine Technologie durch eine Firma zum ersten Mal angewendet. Ist das System eben dieser Sektor, dann handelt es sich in der Tat um eine Innovation. Das gilt selbst dann, wenn diese Innovation aus einem anderen Sektor „zugewandert“ ist, also auf der Ebene der gesamten Volkswirtschaft keine Innovation im Sinne von etwas völlig Neuem darstellt. Ist der Bezugsrahmen noch enger zum Beispiel begrenzt auf eine Firma, dann kann etwas was „neu“ ist für diese Firma, durchaus bereits anderer Stelle erprobt sein. Die Definition von dem Neuen hängt also entscheidend davon ab, was das zu beobachtende System ist. Gehört das Neue noch nicht zu dem Verhaltens- oder Möglichkeitspool von eben diesem System, dann sprechen wir von einer Innovation, die sich durchsetzen kann – oder auch nicht, und die zumindestens das Potential einer Destabilisierung des bestehenden Zustandes in sich trägt.

Ein solches Herangehen bietet die Möglichkeit, verschiedene Erneuerungsprozesse zu klassifizieren, sie mit Elementarprozessen der Aneignung auf der Mikroebene zu verbinden und gleichzeitig in ihrer möglichen Auswirkung auf die Systemstruktur auf der Makroebene zu untersuchen. Jede mathematische Modellierung erfordert eine Formalisierung der möglichen Prozesse und Zustände. Die dabei notwendige Reduktion erzwingt auch eine gewissen Klarheit. Der Gewinn liegt in der Möglichkeit, zu mindestens für die dynamischem Modelle über die wir in diesem Kapitel sprechen, prinzipielle Aussagen über Systementwicklung zu machen, Szenarien zu testen, und nach empirischen Belegen zu suchen – auf der Ebene der Phänomene ebenso wie auf der Ebene der Prozesse, die sie erklären.

# 3.3. Dynamik sozialer und kognitiver Suchprozesse auf komplexen Landschaften – geometrisch orientierte Modelle von Innovationsprozessen

## 3.3.1 Beschreibung von Gruppen bzw. Populationen im Merkmalsraum

Der klassische populationsdynamischen Zugang behandelt die Einheiten der Evolution als klassifizierbar, voneinander unterscheidbar und damit abzählbar. Jeder Einheit (Verhaltensmöglichkeit, Sorte, Gruppe, Population) wird eine Nummer, d.h. eine natürliche Zahl $i = 1,2,3,...$ zugeordnet. Die Populationen, oder allgemein im Falle sozialer Systeme die Gruppen, bilden somit eine abzählbare Menge. Jede Einheit wird durch eine quantifizierbare, zeitabhängige Grösse – eine reelle Zahl $x_i(t)$ – charakterisiert. Das sind im Fall der evolutionären Ökologie die Dichte bzw. Anzahl der Individuen konkurrierender Arten (etwa in Räuber- und Beutesystemen), in der Theorie der molekularen Evolution chemische Konzentrationen verschiedener makromolekularer Sorten, im Fall von sozial-technologische Systemen die Zahl der Nutzer einer Technologie, im Fall des Wissenschaftssystems die Wissenschaftler, die auf einem gestimmten Gebiet arbeiten, und im Fall von Lernprozessen in Gruppen können es die Anhänger einer bestimmten Problemlösung sein. Im Fall der Kompetenzmodellierung sind es Individuen, etwa in einem Unternehmen, die ein ähnliches Profil von Kompetenzen besitzen bzw. anwenden.
Die Synergetik bezeichnet diese quantitativen Grössen $x_i(t)$ als Ordner, die die Populationen auf der Makroebene repräsentieren. Die Zeitabhängigkeit der Ordner wird durch eine mathematische Abbildung, in der Regel dargestellt durch gewöhnliche Differentialgleichungen, definiert. Damit wird in einem gewissen Sinne von der Merkmalsstruktur der einzelnen Individuen in der Population und deren Veränderung abstrahiert und die Konkurrenz zwischen verschiedenen Populationen ins Zentrum gestellt. In der mathematischen Beschreibung führt dieser Ansatz zu Systemen nichtlinearer gewöhnlicher Differentialgleichungen.
Dem Einwand, dass bei diesem reduktiven Vorgehen gewissermassen alle Individuen als gleich angesehen werden, lässt sich dadurch begegnen, dass man von der deterministischen Beschreibung in der Sprache der Differentialgleichungen zu einer stochastischen Beschreibung mit Mastergleichungen übergeht (siehe Weidlich, Haag, 1983; Ebeling, Feistel 1982, 1986; Saam 1995).

**Tabelle 2: Verschiedene Klassen mathematischer Modellbildung (Ebeling, Scharnhorst, 1999)**

| | Diskrete Modelle | Kontinuierliche Modelle |
|---|---|---|
| Deterministisch | $\frac{d}{dt}x_i = f_i(x_1,...x_i,...x_n;\vec{u})$ <br> $\vec{u} = \{u_1,u_2,...,u_p\}$ <br> *Systemparameter* | |
| | Nicht-lineare gewöhnliche Differentialgleichungssysteme, z.B. Lotka-Volterra Systeme | $\partial_t x(\vec{q},t) = f(x(\vec{q},t),U)$<br>$U = \begin{Bmatrix} u_1(\vec{q}),\ldots,u_i(\vec{q}), \\ u_j(\vec{q};x(\vec{q},t)),\ldots \end{Bmatrix}$<br>*Parameterfunktionen oder -funktionale*<br><br>Partielle Differentialgleichungen, z.B. Reaktions-Diffusions-Gleichungen |
| Stochastisch | $\frac{d}{dt}x_i = f_i(x_1,\ldots x_i,\ldots x_n;\vec{u}) + F_i(t)$<br>$F_i \ldots stochastische\ Quelle$<br><br>Langevin oder Fokker-Planck Gleichungen | Funktionale Fokker-Planck Gleichungen |
| | $N_i = 1,2,\ldots \quad x_i = N_i / N$<br>$P(N_1,\ldots N_i,N_j,\ldots)$<br>$\partial_t P = WP$<br>Mastergleichung | Funktionale Mastergleichung |

Diskrete Modelle, sowohl deterministische als auch stochastische, gehören inzwischen zu dem Grundbestand mathematischer Modellierung von sozialen Phänomenen. Weniger Aufmerksamkeit wird dagegen den so-genannten landschaftsorientierten Modellen gewidmet. Zugegebenermaßen ist deren mathematische Komplexität höher, und die empirische Validierung schwieriger. Andererseits hat dieser mathematische Ansatz Potentiale, die der Spezifik sozialer Selbstorganisationsprozesse sehr entgegenkommen. Zu dieser Spezifik gehört:

- Verhaltensmuster und Verhaltensmöglichkeiten von Individuen lassen sich nur schwer einer diskreten Typologie zuordnen. Meistens sind Verhaltensmuster in ein Kontinuum von Verhaltensmöglichkeiten eingebettet, die ineinander übergehen.
- Für soziale Prozesse gilt ebenso wie für andere, natürliche Prozesse der Selbstorganisation, das die auf der Makroebene gebildeten Strukturen, als Rahmenbedingen, den individuellen Verhaltensspielraum einschränken, aber die Bewertung der Strukturen, was als jeweils erfolgreich oder nicht angesehen wird, ist stark einem sozialen Prozess des Unterhandelns unterworfen. Externe Kriterien spielen dabei eine Rolle, aber noch mehr wird das Heutige an dem bereits Existierendem gemessen.

Die Nachteile dieser komplexeren Beschreibung sind, wie bereist erwähnt:

- Der Merkmalsraum, in denen die Gruppen angeordnet sind, ist nicht einfach zu definieren. Die für das Wissenschaftssystem gewählte Landschaft (Abbildung 7) suggeriert einen zwei-dimensionalen Raum, stellt aber eine zwei-dimensionale Projektion eines höher-dimensionalen Netzwerkes von Beziehungen dar. Die Interpretation und Metrisierung der zwei Dimensionen ist nicht trivial.
- Noch schwieriger ist es die Bewertungsfunktion empirisch nachzuweisen, die die Dynamik in diesem Merkmalsraum steuern.

Arbeiten, die konzeptionell den Landschaftsmodellen am nächsten stehen, finden sich in der so-genannten *computational philosophy,* einem Zweig der Wissensphilosophie und Erkenntnistheorie. In diesem Bereich sprechen Autoren von sogenannten erkenntnistheoretischen Landschaften, allerdings ohne bisher eine direkten Bezug zur Empirie herzustellen. (Payette, 2012) Es scheint uns daher angemessen, diesem vielversprechenden, aber noch nicht in Blüte stehendem Theorie-und Methodenrahmen, hier nochmal besondere Aufmerksamkeit zu schenken.

### 3.3.2. Evolutionäre Dynamik im Merkmalsraum

In diesem Abschnitt stellen wir den kontinuierlichen Ansatz vor und fassen wesentliche Züge des Modells kurz zusammen (Tabelle 2, grau unterlegte Zelle). Wir beschreiben eine Population durch Merkmale (Feistel, Ebeling 1989), die durch einen reell-wertigen Satz von *d* Variablen $\vec{q} = \{q_1, q_2, ..., q_d\}$ charakterisiert werden. Damit wird ein abstrakter Merkmalsraum ***Q*** definiert. Die einzelnen Merkmale $q_i$ bilden die Koordinaten von ***Q*** der die Dimension *d* hat. In der Regel gehen wir davon aus, dass in sozialen Systemen viele differenzierende Merkmale vorliegen, und daher *d* eine grosse Zahl ist. Die Koordinaten $q_1,...,q_d$ kennzeichnen die Ausprägung des jeweiligen Merkmals. Ein Punkt im Raum ***Q*** kennzeichnet somit den aktuellen Zustand eines Individuums durch seine Merkmalskonfiguration. Wenn wir als Beispiel einen abstrakten Problemlösungsprozess in der Wissenschaft betrachten, dann lassen sich theoretische, methodische und empirische Merkmale definieren. Diese können u.U. dadurch gemessen werden, dass man die publizierten Arbeiten des Wissenschaftlers verschiedenen Gegenstandsbereichen (im Sinne einer wissenschaftlichen Klassifikation) oder verschiedenen konzeptionellen Dimensionen (im Sinne eines Vektorraumes der verwendeten Termini) zuordnet. Die Veränderung realisierter Merkmale führt zu einer Bewegung der entsprechenden Punkte analog zur Bewegung von Teilchen anhand zeitlich variierender Koordinaten im Ortsraum. Für das Beispiel des Wissenschaftssystems, stellt die Bewegung der Punkte den Suchprozess nach neuen Fragestellungen dar. Dieser lässt empirisch anhand von Publikationen verfolgen. Für einen einzelnen Wissenschaftler lässt sich dieser Wechsel von Themen, auch Feldmobilität genannt (Ebeling, Scharnhorst 2009), als ein spezifisches Strichcodemuster visualisieren. (Hellsten 2007)

Die Bewegung dieser Punkte wird im kontinuierlichen Modell nicht individuell beschrieben, sondern über eine Dichte-Funktion $x(\vec{q},t)$. Die Dichtefunktion ist eine reellwertige nichtnegative Funktion über dem Raum ***Q***, die eine komplizierte Struktur mit vielen Maxima besitzt. Die meisten Stellen im ***Q***-Raum sind unbesetzt, d.h. die Dichtefunktion ist dort Null und nur an wenigen ,,günstigen" Stellen liegen lokale Maxima der Dichtefunktion vor. Die Dichtefunktion $x(\vec{q},t)$ tritt an die Stelle der diskreten Funktion $x_i(t)$. Populationen werden aus Gruppen von Elementen/Individuen mit ähnlichen Eigenschaften gebildet, die räumlich einander benachbart sind. Eine Population entspricht geometrisch einem lokalen Maximum der Dichtefunktion. Die räumliche Anordnung der Populationen entspricht der Nähe oder Distanz ihrer Merkmalsstruktur. Man kann auch sagen, dass in einem solchen Merkmalsraum Merkmalskombinationen durch bestimmte Orte, Individuen durch realisierte oder besiedelte Orte, Populationen als lokale Anhäufungen solcher besiedelten Orte und Dynamik durch die zeitlichen Verschiebungen der lokalen Maxima charakterisiert sind.

Die Regeln der zeitlichen Veränderung der Besiedlung definieren die Systemdynamik. Diese werden durch eine partielle Differentialgleichung für die Funktion $x(\vec{q},t)$ beschrieben.

Ein wesentliches Merkmal einer Evolutionsdynamik, ist die Annahme der Existenz einer Bewertungsfunktion über dem Merkmalsraum. Eine solche Bewertung – im einfachsten Fall durch eine skalare Funktion $w(\vec{q},t)$ gegeben – ordnet jedem Raumpunkt einen Funktionswert zu der ein Maß für die lokale Fitness ist. Die so eingeführte Bewertungsfunktion bildet eine Landschaft über dem abstrakten Merkmalsraum. Im Endeffekt beschreibt die Evolutionsdynamik die Wechselwirkung von zwei Funktionen, von zwei Landschaften im Merkmalsraum. Es handelt sich dabei um die Besetzungsfunktion, die eine Realisierung bestimmter Merkmale darstellt, und in der Regel einer empirischen Bestimmung leichter zugänglich ist; und die Bewertungsfunktion – oder Fitness, die Veränderung der Besetzungsfunktion zu erklären vermag, aber häufig nur indirekt empirisch nahgewiesen werden kann.

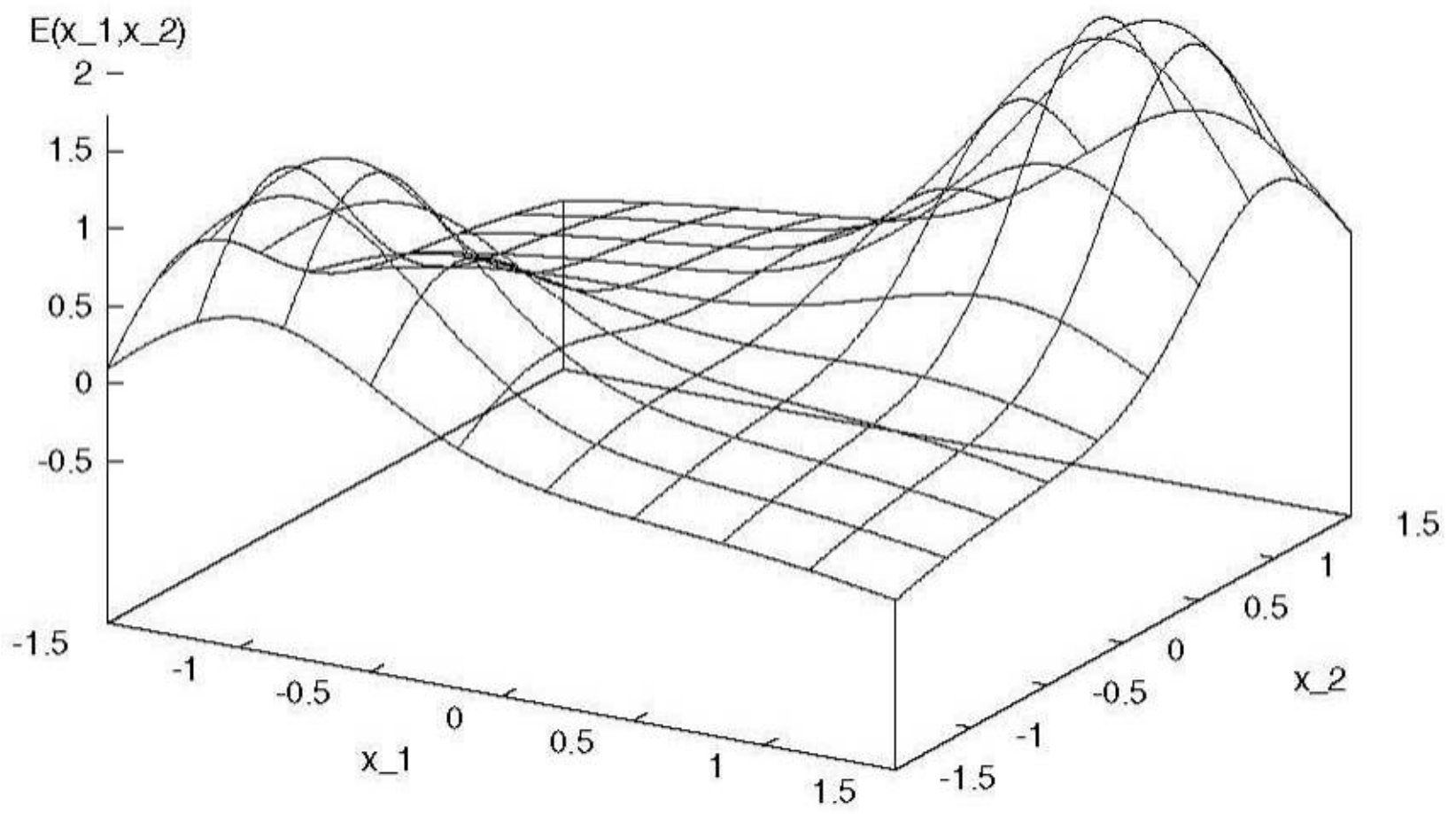

**Abbildung 9 Beispiel einer Fitnesslandschaft (Bewertungsfunktion) mit zwei Gipfeln**

Die Fitnessfunktion ist eine in der Regel **unbekannte** Landschaft. Diesem Problem der Definition und Bestimmung einer Fitnessfunktion wird dadurch Rechnung getragen, dass diese als Zufallslandschaft mit bestimmten statistischen Eigenschaften dargestellt wird. Damit lassen sich Verbindungen zur Physik ungeordneter Systeme (Anderson 1978) herstellen.

Die Bewertungsfunktion $w(\vec{q},t)$ ihrerseits kann verschiedene Gestalt annehmen. Im einfachsten Fall hängt diese nur von den Merkmalen und ist zeitlich konstant. Sozialer Dynamik angemessener ist eine Bewertungsfunktion, die sich in der Zeit verändert, so wie sich Werte und Normen in sozialen Systemen immer wieder ändern können. Eine spezifische Beschreibung dieser Zeitabhängigkeit sind sogenannte Lotka-Volterra-Systeme. Dabei hängt die Bewertungsfunktion explizit von der Besiedlung ab, und verändert ihre Gestalt, je nachdem welcher Bereich im Merkmalsraum besiedelt wird $w(\vec{q};x(\vec{q},t))$.

Den universellen Eigenschaften evolutionärer Prozessen folgend, ist Dynamik mit den Reproduktionseigenschaften bestimmter Merkmalskombinationen verbunden. Für eine allgemeine Evolutionsdynamik vom Darwin-Typ (Replikationsansatz) gilt: (Fesitel, Ebeling 1982; Feistel, Ebeling 1984; Feistel, Ebeling 1990):

$$\partial_t x(\vec{q},t) = w(\vec{q},\{x\})x(\vec{q},t) + Mx(\vec{q},t) \tag{1}$$

dabei ist $w(\vec{q};x(\vec{q},t))$ die verallgemeinerte Fitnessfunktion. Sie beschreibt den Selektionsaspekt der evolutionären Suchdynamik. Hängt *w* nur von den Merkmalen $\vec{q}$ ab, so handelt es sich um eine stationäre Landschaft. Im einfachsten Fall gilt

$$w(\vec{q};x(\vec{q},t)) = E(\vec{q}) - \langle E \rangle \tag{2}$$

Hierbei bezeichnen die eckigen Klammern <*E*> den sogenannten sozialen Durchschnitt (gemittelt über die Population):

$$\langle E \rangle = \int E(\vec{q}\,')x(\vec{q}\,',t)d\vec{q}\,'.$$

Im (komplizierteren) Fall eines Fitnessfunktionals hängt die Bewertung eines Ortes $\vec{q}$ auch von der Besiedlung $x(\vec{q},t)$ im gesamten Merkmalsraum der Kompetenzen ab. Eine Veränderung dieser Besiedlung führt auch zu einer Veränderung der Bewertung, und eine Ko-Evolution zwischen Fitnessfunktion und Populationsdichte liegt vor. Ein Beispiel für eine solche Kopplung stellt der folgende Lotka-Volterra-Ansatz dar:

$$w(\vec{q};x(\vec{q},t)) = a(\vec{q}) + \int b(\vec{q},\vec{q}\,')x(\vec{q}\,',t)d\vec{q}\,' \tag{3}$$

In diesem Fall setzt sich die Bewertung einer Merkmalskombination aus zwei Anteilen zusammen. Der erste Term in Gleichung (3) $a(\vec{q})$ stellt eine Bewertung der reproduktiven Aspekte der Merkmale dar. Der zweite Term beschreibt die Wechselwirkungen der Merkmale, d.h. den – über das ganze Raumgebiet integrierten und durch die Koeffizienten $b(\vec{q},\vec{q}\,')$ gewichteten – Einfluss anderer besiedelter Orte. In diesem Fall spricht man von einer adaptiven Landschaft (Conrad 1978; Conrad, Ebeling 1992).

Im Spezialfall eines Delta-Funktions-Kernes kann man Gl. (2) aus Gl. (3) zurückerhalten. Die evolutionäre Dynamik wird dann als ein Suchprozess in einer sich ständig verändernden adaptiven Fitnesslandschaft verstanden. Es wurde gezeigt, dass diese Art der Modellbeschreibung von spezifischer Relevanz für sozio-technologische Systeme ist. (Scharnhorst, 2001) In diesen bestimmen nichtlineare Rückkopplungen zwischen dem Handeln der Akteure auf der individuellen Ebene und Koordinations- und Bewertungsprozessen auf makroskopischer Ebene wesentlich die Systemdynamik.
Neben der Bewertungsfunktion ist die Grösse *M*, der sogenannte Mutationsoperator in Gleichung (1) entscheidend für die Evolutionsdynamik. Dieser Term beschreibt den Elementarprozess der Innovation auf der Mikroebene, nämlich der Besiedlung von Problemstellungen mit Merkmalen, die bisher nicht realisiert wurden. Ob diese *elementaren Innovationsakte* auch zu einer globalen Systemveränderung[6] führen, lässt sich nicht von vornherein entscheiden. In einer ersten Näherung kann man die elementaren Suchprozesse innerhalb und am Rande der um bestimmte Zentren konzentrierten Populationen in einer ersten Näherung als diffusionsartig beschreiben

$$Mx(\vec{q},t) = D\Delta x(\vec{q},t) \qquad (4).$$

Die Wirkung der Evolutionsdynamik lässt sich folgendermaßen veranschaulichen. Mutation oder spezifischer Diffusion führt zu einer Ausdehnung des besiedelten Teils des Raumes. Die Diversität der im System realisierten Merkmale nimmt zu; die Besiedlung verteilt sich auf größere Bereichen. In anderen Worten, wir beobachten eine Koexistenz von vielen verschiedenen Merkmalskombinationen. Gleichzeitig führt die Existenz des ersten Terms in Gleichung (1), des Teils der Dynamik, der an die Bewertungsfunktion gekoppelt ist, dazu dass nicht alle diese einmal realisierten Kombinationen auch im gleichen Maße wachsen. Da Ressourcen in der Regel begrenzt, im Fall des Wissenschaftssystem etwa die Zahl der Wissenschaftler beziehungsweise die Gelder für Forschung, kommt es zwangsläufig zu einem Wettbewerb um diese Ressourcen. Die erste Term in Gleichung (1) führt dazu, dass sich die Besiedlung um die Maxima der Bewertungsfunktion zusammenzieht.
In der Regel geht die Erkundung neuer, konkurrenzfähiger Bereiche (Problemlösungen) von den Rändern der bestehenden Besiedlungen mittels eines undifferenzierten Suchprozesses aus. In der kontinuierlichen Beschreibung definiert die Systemdynamik (Gl. (1) mit (2) und (4)) im einfachsten Fall eine zeitliche Variation der global definierten Besetzungsfunktion $x(\vec{q},t)$ durch die partielle Differentialgleichung (Darwin-Dynamik, Darwin-Strategie)

$$\frac{\partial x(\vec{q},t)}{\partial t} = x(\vec{q},t)\left[E(\vec{q}) - \langle E \rangle\right] + D\Delta x(\vec{q},t) \qquad (5)$$

Die sich herausbildenden Zentren der Verteilungsfunktion $x(\vec{q},t)$folgen teilweise den Maxima der Bewertungsfunktion. Die Dynamik hängt sehr stark (Feistel, Ebeling 1990) von den Annahmen über die Struktur der Fitnessfunktion ab. Einmal gebildet, kann man die (Lokalisations-)Zentren der Populationswolken als Analogon zu den diskreten Typen (als Ordner des Systems) behandeln.

[6] etwa im Sinne des Wechsels der Besiedlung von einem Maxima der Bewertungsfunktion zu einem anderen.

Die Vorteile eines kontinuierlichen Ansatzes liegen auf der Hand. In einem diskreten Modell entsteht immer die Frage, wie etwas neues, eine neue Gruppe oder Dimension im Modell erfasst werden kann. Die Ebeling'sche Schule hat dazu den Begriff des *underoccupied,* des unterbesiedelten Systems eingeführt. (Ebeling, Sonntag 1986) Dabei handelt es sich um einen gewissermaßen mathematischen Trick. Wir nehmen an, dass die Anzahl der möglichen Gruppen oder Verhaltensmöglichkeiten *n* gross ist, und unter diesen immer einen genügende Anzahl, noch nicht realisierte – „leere" Gruppen vorhanden ist, die im Laufe der Evolutionsdynamik besiedelt werden können.
Im kontinuierlichen Modell sind solche Annahmen nicht nötig. Die Beschreibung der Besiedlung neuer Bereiche ist integrierter Bestandteil der Dynamik. Aber auch das Verschmelzen von Gruppen, das Aufspalten einer Gruppe in mehrere, kurzum alle Prozesse, die in einem diskreten Modell einer Veränderung der Taxonomie entsprechen würden, sind Teil der Systemdynamik. Wie bereits am Eingang dieses Abschnittes erwähnt, sind diese Vorteile der Beschreibung mit einer wachsenden Komplexität verbunden.
Das im Abschnitt 3.2. Kernproblem des Entstehens und die Verbreitung von Innovationen lässt sich im kontinuierlichen Modell auf die Frage abbilden wie Individuen von einem Ort einer lokal maximalen Bewertung zu einem anderen Ort mit einer, möglicherweise global höheren maximalen Bewertung überwechseln können.
Für eine relativ einfache Dynamik sind dabei auch noch analytische Aussagen möglich. Wie wir anderen Stelle im Detail diskutiert haben (Scharnhorst, Ebeling 2005), lassen sich folgende Zusammenhänge von Mikrodynamik und Strukturveränderung auf der Makroebene erklären, und Hypothesen zu Konsequenzen bestimmter Interaktionsmuster für das Gesamtsystem formulieren:
Der Mechanismus der Bewertungsfunktion beinhaltet einen Vergleich zwischen verschiedenen Merkmalskombinationen. Das lässt sich auch so auffassen, dass die Individuen selbst sich, d.h. ihre Position im Merkmalsraum und deren Bewertung vergleichen. Ausgehend von diesem Vergleich kann das Individuum sich dann verschieden verhalten. Es kann die Merkmalskombinationen anstreben, die bereits besiedelt und als besser erkannt sind. Es kann sich auch an der reinen Häufigkeit der Besiedlung bestimmter Bereiche (unabhängig von einer Bewertung) orientieren. Je nach dem, welches Verhalten – im Mittel von den Individuen bevorzugt wird, verändert sich die Wahrscheinlichkeit ein anderen Maximum zu finden, und die Zeit in der ein solcher Wechsel vollzogen wird. Die Dynamik lässt sich auch noch feiner analysieren. Herrscht in einer Gruppe etwa eine starke Neigung sich auf einander zu beziehen, und dass eigene Verhalten von der Position anderer bestimmen zu lassen, dann wird es länger dauern bis eine Alternative „entdeckt" und besiedelt wird. Denn für diesen Akt ist es notwendig, dass einige Gruppenmitglieder sich aus diesem Gruppenzwang lösen, zeitliche Verschlechterungen akzeptieren, und bereit sind das Risiko einzugehen scheinbar blind anderen Merkmalskombinationen auszuprobieren. Besteht aber in einer Gruppe ein solcher Gruppenzwang **und** haben einige Individuen die Alternative gefunden, dann wird die Gruppe viel schneller folgen, als in einer Situation, in der der Zusammenhalt der Gruppe viel loser ist.

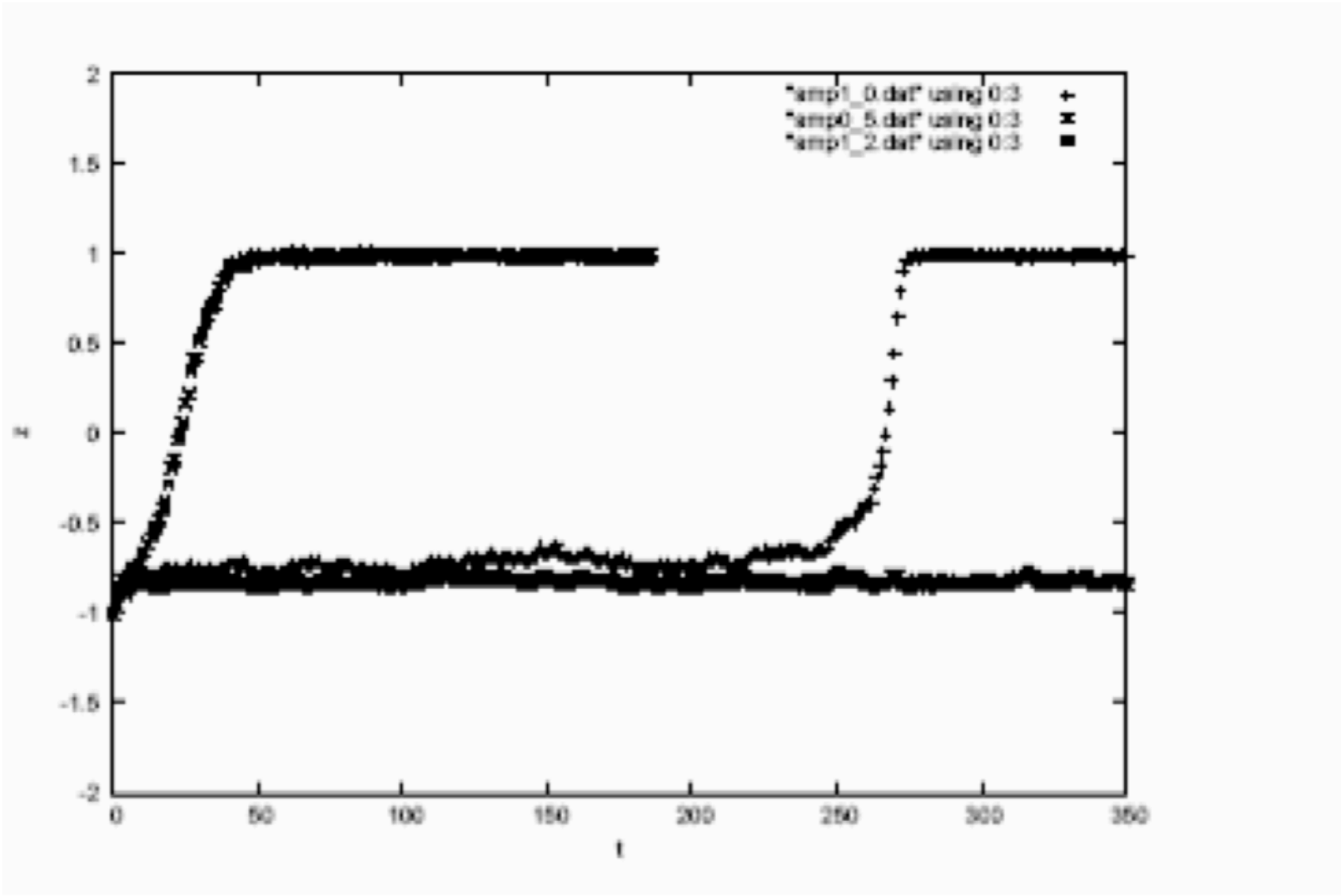

**Abbildung 10: Der Zeitverlauf von Übergängen einer Gruppe zwischen zwei Maxima. Die drei Kurven representieren die Positions eines gedachtes Mittel- oder Schwerpunktes der Gruppe. Die Suche beginnt jeweils bei der Position (-1). An der Zeitachse lässt sich ablesen, wann der Übergang stattfinden, wenn überhaupt, und an der Steilheit der Kurve, wie schnell dieser Übergang passiert. Die drei Kurven entsprechen, von links nach rechts, einer zunehmenden Interaktion oder Gruppenzwang.**

## 3.5. Evolino – ein evolutionäres Suchspiel

Der Vorteil von einer mathematischen Modellierung liegt nicht nur in analytischen Resultaten. Modelle – und vor allem komplexe Modelle - können auch mit Hilfe numerischer Simulationen erkundet werden. Diese Simulationen lassen sich verschieden gestalten: als methodische Erkundung von Systemverhalten; als interaktive Plattformen mit denen Nichtspezialisten an die Gedankenwelt eines Modell herangeführt werden; oder als Spiele, deren Zusammenhang von der mathematischen Modellbildung nur noch metaphorisch ist, die aber dennoch zur einer anderen Perspektive auf bestimmte Phänomene anregen können. (Beaulieu et al. 2013)
Wir wollen in diesem Abschnitt ein Spiel auf einer Suchlandschaft darstellen – EVOLINO (Scharnhorst, Ebeling 2005), in dem die oben dargestellten Prinzipien einer Evolutionsdynamik auf einer Landschaft realisiert sind. Ein Anliegen bei der Programmierung von EVOLINO war, eine interaktive Simulationen zu schaffen, die von einem mathematisch nicht besonders vorgebildeten Nutzer bedient werden kann und dennoch das mathematische Modell und nicht nur die Konzepte dahinter verkörpern. (Hüsing und Scharnhorst, 2007, 2013).

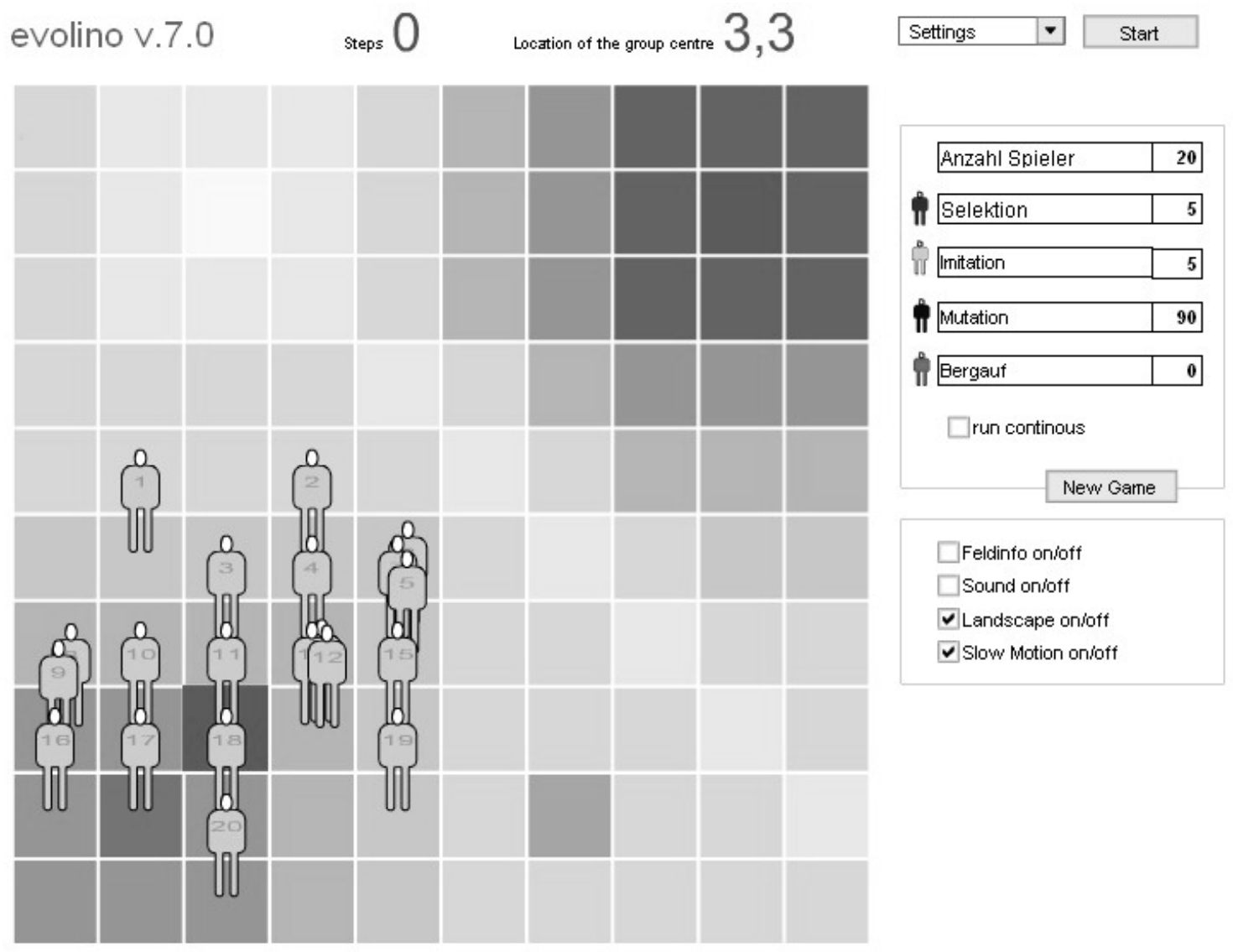

**Abbildung 11: Startbild von *Evolino* (Version 7.0). 20 Spieler (Agenten) werden zufällig in der linken unteren Hälfte des Spielfeldes verteilt. Die Parameter für Selektion, Imitation und Mutation können eingegeben werden. Der Mittelpunkt der Gruppe wird durch ein orange farbenes Quadrat angezeigt. Dessen Koordinaten oben in der Mitte angegeben werden. Eine Anzeige „Steps“ zählt die Spielschritte.**

In seiner allgemeinen Form ist *Evolino* ein Evolutionsspiel von Agenten bzw. Individuen in einem Spielraum (Spielfeld=Merkmalsraum). Das Spielfeld von 10x10 Kästchen ist eine Diskretisierung des bisher kontinuierlichen Merkmalsraums. Über diesem Spielfeld ist eine Landschaft definiert. Diese wird durch die unterschiedliche Farbe der Kästchen markiert. Die Tiefe des Blautons entspricht dabei der Höhe der Landschaft. Ein Schalter (Feldinfo on/off) ermöglicht es, sich diese Werte auch anzeigen zu lassen. Die Landschaft besteht aus zwei Gipfeln und einem Zwischenmaxima. Ziel der Individuen ist es das neue Maxima im rechten oberen Quadranten zu finden. Dabei ist den Individuen die Landschaft nur an den Stellen bekannt, an denen sie sich befinden.

Das Spiel verwendet drei evolutionäre Mechanismen: Selektion, Imitation und Mutation (als dem Wechsel eines Individuums zu einer rein zufälligen Nachbarposition. Diese Mechanismen werden als Verhaltensregeln für die Spieler festgelegt. Durch die Parametereingabe wird festgelegt mit welcher Häufigkeit welcher Evolutionsmechanismus eingesetzt wird. Alle Parameter addieren sich zu hundert. Eine Parameterkombination von (Selektion, Imitation, Mutation) = (10, 10, 80) bedeutet, dass auf hundert Evolutionsschritte, 10 Selektionsschritte, 10 Imitationsschritte und 80 Mutationsschritte kommen. Programmiertechnisch ist die Mutation der Grundschritt, der durch Selektion und Imitation unterbrochen wird.

SELEKTION bedeutet, dass zwei Spieler zufällig ausgewählt werden, dann wird verglichen wie hoch (Güte) die Landschaft an ihren Positionen ist, der Spieler von der schlechteren Position geht zu dem Spieler auf der besseren Position über. Haben beide gleich gute Positionen erfolgt eine Aktion. Durch diesen Prozess können sich Spieler bewusst verbessern, d.h. höher gelegene Teile der Landschaft besiedeln.

IMITATION bedeutet, dass zwei Spieler zufällig ausgewählt werden, dann wird verglichen, wie stark ihre Positionen besiedelt sind ist, der Spieler von der weniger besiedelten Position geht zu dem Spieler auf der stärker besiedelten Position über. Sind beide Positionen gleich stark besetzt erfolgt keine Aktion. Mit dieser Regel wird soziale

Imitation als Nachahmung positiv besetzt. Viele Spieler auf einem Feld üben eine Anziehungskraft aus.
MUTATION bedeutet, dass ein Spieler zufällig ausgewählt wird, und dann zufällig ein Nachbarfeld (oben, unter, links oder rechts) zu seiner Position ausgewählt wird, auf das er sich bewegt. Mit dieser Regel wird die Erkundung der Landschaft als lokaler Prozess modelliert. Die Mutation ist die einzige Regel mit der neue Bereiche der Landschaft erkundet werden können (Innovation).
Als eine Randbedingung bei dem begrenzten Spielfeld wurden sogenannte reflektierende Wände angenommen, d.h. würden Bewegungen aus dem Spielfeld herausführen, werden die Spieler wieder zurückgesetzt. Mit den obigen Regeln wurde ein Spiel geschaffen, dass bei der evolutionäre Suche die selben Eigenschaften zeigt wie das kontinuierliche Modell. Es handelt sich dabei um eine Art zellulärer Automat, der sich von üblichen zellulären Automaten dadurch unterscheidet, dass er auch nichtlokale Prozesse bzw. Wechselwirkungen enthält (Selektion und Imitation) und dass eine Landschaft als zusätzliches Element über dem Raum definiert ist.
Der Parameterraum von *Evolino* ist gross genug um verschiedene Spielverläufe zu erkunden. Im allgemeinen lässt sich sagen, dass eine relativ hohe Selektion dazu führt, dass die Gruppe immer wieder um die bereits erreichten Gipfel in der Landschaft konzentriert wird. „Ausreisser“, die den Weg zu dem nächst höheren Gipfel durch das Tal wagen, werden relativ schnell wieder eingefangen.

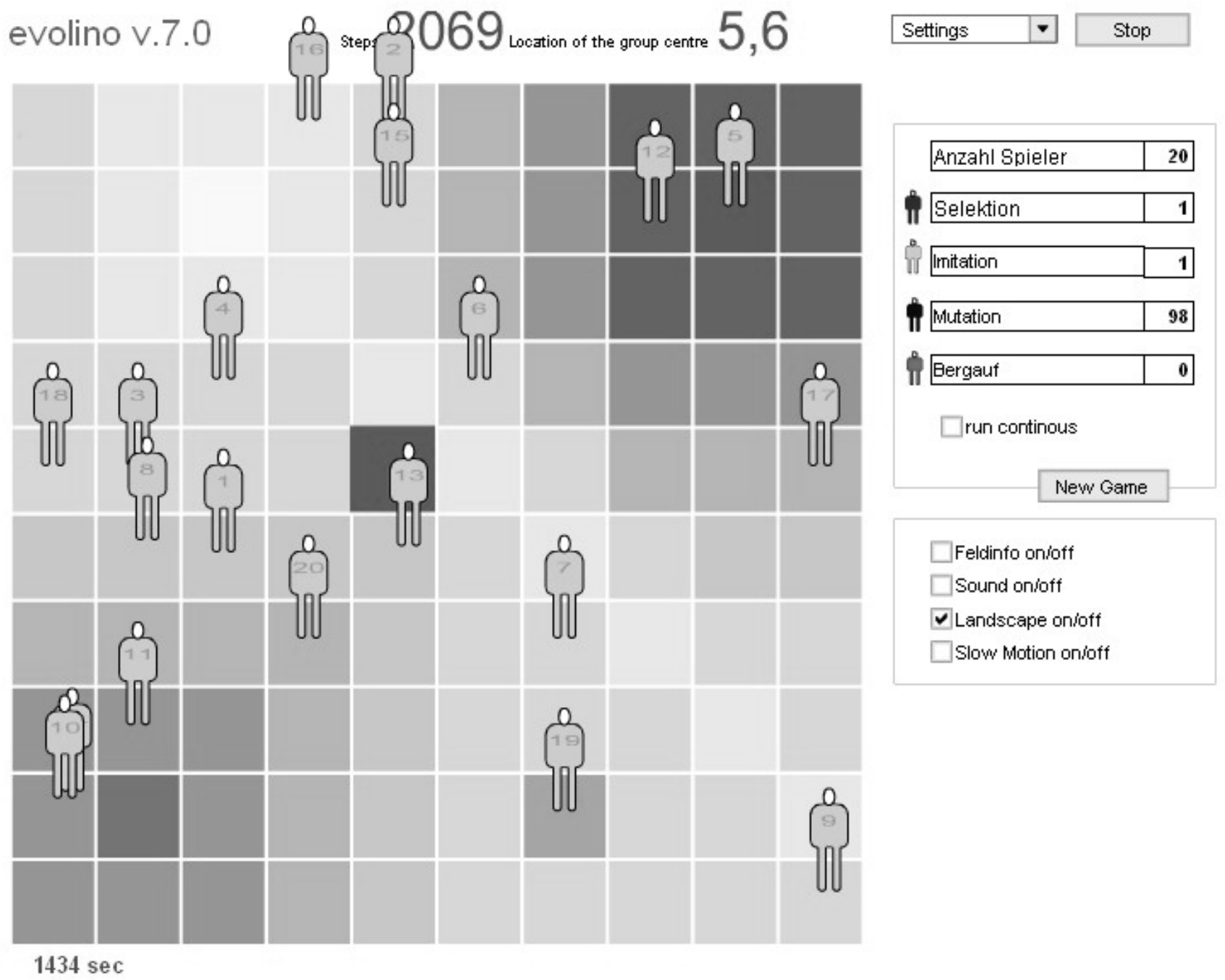

**Abbildung 12: Situationsbild eines Spiels mit relativ hoher Mutationsrate (98) nach ca. 2000 Schritten. Die Dispersion ist sehr hoch, das` ganze Suchfeld wird erkundet. Aber der (rote) Schwerpunkt liegt noch in der Mitte des Spielfeldes und nur zwei Spieler haben bisher den Zielbereich Ziel gefunden. Duch Absenken der Mutationsrate und Vergrosserung der Selektionsrate kann man erreichen, dass eine grossere Zahl von Spielern und damit auch der Schwerpunkt den Zielbereich finden.**

Das interaktive Spiel „Evolino“ ist im Internet verfügbar (Huesing und Scharnhorst, 2007, 2013) Es liegt sowohl in einer allgemeinen Form vor, als auch angewendet auf Prozesse des Wissensmanagements, und lebenslangen Lernens. Die letztere Anwendung wurde im Rahmen eines Forschungsverbundes entwickelt (Erpenbeck et al. 2006), wobei das Anliegen war, eine Verbindung zwischen empirischer Erhebung von Kompetenzen, den selbstorganisierenden, gruppendynamischen Prozessen in Problemlösungssituationen, und mathematischen, formalen Theorien der Selbstorganisation herzustellen. Der Kontext dieses Projektes lag in früheren Forschungen zu einer Benutzung von Ideen der Selbstorganisation für Problem von Management und Organisationen (Beer 1968, Laszlo et al. 1992). Häufig werden in diesem Bereich Konzepte und Ideen der Selbstorganisation eher assoziativ eingesetzt. Mit dem Spiel Evolino haben wir versucht, den Bezug zu mathematischen Theorien zu erhalten und dennoch die Schwelle des Zugangs auch für Experten ausserhalb der Selbstorganisationstheorien zu senken.

Die Erfahrungen mit diesem Projekt wurde aus wissenschaftstheoretischer Sicht an anderer Stelle beleuchtet. (Beaulieu et al. 2013) Zusammenfassend aber lässt sich feststellen, dass die sich Erklärungskraft komplexer mathematischer Modellbildung nicht ohne Bedeutungsverluste in einen Bereich ausserhalb der mathematischen Sprache übertragen lässt. Die Skylla und Charybdis der Übertragung von mathematischen Modellen, die in anderen Wissenschaftsbereichen entwickelt wurden, in die Sozialwissenschaft sind eine zu komplexe Mathematik auf der einen Seite, die den Verbreitungsbereich stark einschränkt, und eine zu grosse Verallgemeinerung auf der anderen Seite, die Gefahr läuft allgemein Bekanntes zu wiederholen.

## 4. Zusammenfassung

Dieser Beitrag beleuchtet Theorien der Selbstorganisation aus der Physik und ihre Anwendungen in den Sozialwissenschaften. Wir haben uns entschieden, den Bogen von den Anfängen der Selbstorganisationstheorien in den Naturwissenschaften bis zu hoch spezialisierten Einzelanwendungen in den Sozialwissenschaften zu spannen. Unser Anliegen ist es, durch eine historische Darstellung, ein besseres Verständnis für das Potential, und notwendige Bedingungen einer sinnvollen Anwendung zu schaffen, und somit eine weitere Verbreitung zu fördern.

Die heuristische Kraft mathematischer Modelle kann auf ganz verschiedenen Ebenen wirken:

1. Grundsätzlich gilt, dass alle Modelle (mathematische und nichtmathematische) Formalisierungen darstellen, deren reduktiver Charakter gleichzeitig auch zur Ordnung und Prioritätensetzung zwingt.

2. Mathematische dynamische Modelle sind vor allem auch Gedankenexperimente über mögliche Prozesse und deren Wirkungen. Sie können neue Perspektiven auf Phänomene befördern. In dieser Eigenschaft sind sie ein Generator für Fragestellungen, vielleicht auch für die Schärfung des Blicks für neue Phänomene. Mit anderen Worten, sie sind eher Problemerschaffer als

Problemlöser.

3. Der Versuch einer empirische Unterbauung mathematischer Modelle kann zu ganz neuen empirischen Fragstücken führen, auch wenn der herkömmlichen Zyklus von Datenerhebung und Beobachtung, Modellierung, Validierung, Analyse und Abgleich mit Daten nicht konsequent in allen Teilen durchlaufen werden kann.

4. Klassischerweise erwarten wir von Modellen dann doch, dass sie neue Erklärungen für beobachtbare Phänomene bereitstellen, und dabei eben dem im vorherigen Punkt genannten iterativen Zyklus folgen.

Ein Blick auf die Physikgeschichte zeigt die Komplexität des Werdens der Selbstorganisationstheorien, eine Komplexität, die vor allem in den krummen Wegen wissenschaftlicher Erkenntnis und Erkenntnisfortschritts liegt. Dies gilt für physikalische Theorien ebenso wie für soziale. Der Ausflug in die Geschichte zeigt, dass wir selbst in der Physik, eine Pluralität von Modellansätzen finden. Das nimmt nicht hinweg, dass um eine Konsolidierung von Modellierung sozialer Prozesse gerungen werden muss, und darum verschiedene Modellansätze auf einander zu beziehen. Auch wenn Modelle nicht immer ineinander übersetzt werden können, so kann doch ihr verschiedener Gültigkeitsbereich definiert und besser kommuniziert werden. Die gegenwärtige Forschungslandschaft weißt viele, möglicherweise zu isolierte Strömungen auf.

Inspiriert von Wissenschaftsgeschichte und dem Wechselspiel von Empirie und Theorie für die Physik der Selbstorganisation, haben wir uns letztlich für die Darstellung einer Modellklasse entschieden, die aus unserer Sicht gleichzeitig Herausforderungen in mathematischer Hinsicht enthält, und am ehesten dazu geeignet ist, empirische Phänomene von Innovationsdynamik in Wissenschaft, Ökonomie und Wissensmanagement zu erklären. Das Ziel ist es, in Zukunft komplexe Wissenslandschaften, wie die in Abbildung 13, zu erklären und dazu beizutragen, die Rahmenbedingungen in und außerhalb selbstorganisierender sozialer Prozesse so zu gestalten, dass Innovation und Kreativität gefördert werden.

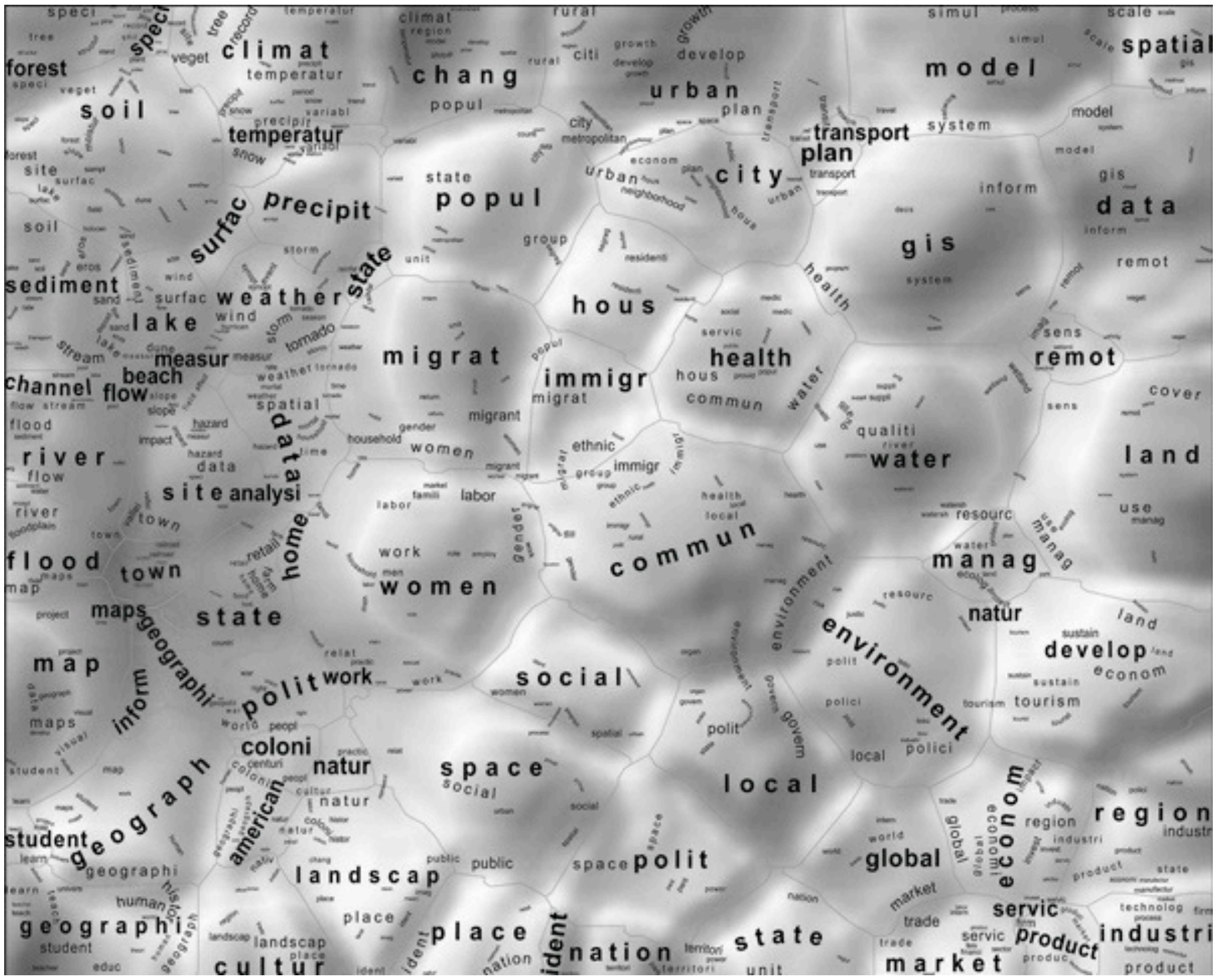

**Abbildung 13: Die Forschungslandschaft der Geologie – André Skupin (2005)**

# 7. Literatur